\documentclass[preprint,12pt]{elsarticle}

 \usepackage{graphicx}

\usepackage{amssymb}
 \usepackage{amsmath}
\usepackage{dcolumn}
\usepackage{bm}
\usepackage{float,color}

\newcounter{bla}

\journal{Computer Physics Communications}

\begin{document}

\begin{frontmatter}



\title{DIRHB -- a relativistic self-consistent mean-field framework for
atomic nuclei}


\author[a]{T. Nik\v{s}i\'{c} \corref{niksic}}
\author[a]{N. Paar}
\author[a]{D. Vretenar}
\author[b]{P. Ring}

\cortext[niksic] {Corresponding author.\\\textit{E-mail address:} tniksic@phy.hr}
\address[a]{University of Zagreb, Faculty of Science, Physics Department}
\address[b]{Physik Department, Technische Universit\"{a}t M\"{u}nchen, D-85747 Garching, Germany}

\begin{abstract}
The {\tt DIRHB} package consists of three Fortran computer codes for the
calculation of the ground-state properties of even-even atomic nuclei using the framework of relativistic
self-consistent mean-field models. Each code corresponds to a particular choice of spatial symmetry:
the {\tt DIRHBS}, {\tt DIRHBZ} and {\tt DIRHBT} codes are used to calculate nuclei with
spherical symmetry, axially symmetric quadrupole deformation, and triaxial quadrupole shapes,
respectively. Reflection symmetry is assumed
in all three cases. The latest relativistic nuclear energy density functionals are implemented in
the codes, thus enabling efficient and accurate calculations over the entire nuclide chart.
%
%
%

\end{abstract}


\end{frontmatter}



{\bf PROGRAM SUMMARY}

\begin{small}
\noindent
{\em Manuscript Title: DIRHB -- a relativistic self-consistent mean-field framework for
atomic nuclei}                                       \\
{\em Authors:}   T. Nik\v{s}i\'{c}, N. Paar, D. Vretenar, P. Ring            \\
{\em Program Title:}  {\tt DIRHB} package (codes {\tt DIRHBS}, {\tt DIRHBZ} and {\tt DIRHBT})\\
{\em Journal Reference:}                                      \\
{\em Catalogue identifier:}                                   \\
{\em Licensing provisions:}   None                                \\
{\em Programming language:}   Fortran 77                                \\
{\em Computer:}  all computers.         \\
{\em Operating system:} all operating systems. The Makefiles are specific
for a Unix OS, and have to be modified for Windows.                                      \\
{\em RAM:} Depends on the imposed symmetry and number of oscillator shells. For
 the triaxial test case it takes 300 Mb.                                          
{\em Classification: 17.22 Nuclear Physics - Hartree Fock calculation}                                         \\
{\em Nature of problem:} Ground-state properties of even-even open-shell nuclei
can be calculated using the framework of self-consistent mean-field models
based on relativistic energy density functionals. The structure of arbitrary heavy nuclei with
spherical symmetry, axially symmetric quadrupole deformation, and triaxial
quadrupole shapes, is modeled using the latest zero- and finite-range
relativistic effective interactions. The particle-particle channel of the effective
inter-nucleon interaction is described by a
separable finite-range pairing force.
\\
   \\
{\em Solution method:} The current implementation of the model computes the mean-field
solution of the nuclear many-body problem for even-even open-shell spherical and quadrupole
deformed nuclei. The codes are used to solve the stationary relativistic Hartree-Bogoliubov equations
in a self-consistent iteration scheme. At each iteration the matrix elements of the equations are
updated using the modified Broyden method or the linear mixing method. The single-nucleon wave functions are 
expanded in a basis of spherical, axially symmetric or triaxial harmonic oscillator, depending on the 
assumed symmetry of the nuclear shape. For calculations that constrain the shape
to specific values of the deformation parameters, the augmented Lagrangian method is used.\\
   \\
{\em Restrictions:} Time-reversal and reflection symmetries are assumed. Open-shell
even-even spherical and quadrupole
deformed nuclei are considered.\\
   \\
{\em Unusual features:} none\\
   \\
   \\
{\em Running time:} Depends on the imposed symmetry and number of oscillator shells. For
 the test cases it runs from few second (spherical) up to few hours (triaxial).\\
   \\

\end{small}

\section{Introduction}

Energy density functionals (EDF) provide an
accurate description of ground-state properties
and collective excitations of atomic nuclei,
from relatively light systems to superheavy nuclei, and from the
valley of $\beta$-stability to the particle drip-lines.
The basic implementation is in terms of
self-consistent mean-field (SCMF) models, in which an EDF is constructed
as a functional of one-body nucleon density matrices that correspond to
a single product state of single-particle or
single-quasiparticle states. This approach is analogous
to Kohn-Sham density functional theory (DFT), that enables
a description of quantum many-body systems in terms of a universal energy
density functional. Nuclear SCMF models
effectively map the many-body problem onto a one-body
problem, and the exact EDF is approximated by simple
functionals of powers and gradients of ground-state nucleon densities and currents,
representing distributions of matter, spins, momentum and kinetic
energy \cite{Bender2003_RMP75-121,Vretenar2005_PR409-101,
Stone2007_PPNP58-587,Niksic2011_PPNP66-519}.

A particular class of SCMF structure models are those based on relativistic (covariant)
energy density functionals. These models have been successfully applied to the
analysis of a variety of nuclear structure phenomena, and the level of accuracy
has reached a level comparable to the non-relativistic
Hartree-Fock-Bogoliubov approach based on Skyrme
functionals or Gogny effective interactions. Here we include the program package DIRHB for
the solution of the stationary relativistic Hartree-Bogoliubov equations for even-even open-shell nuclei
with spherical symmetry, axially symmetric quadrupole deformation, and triaxial quadrupole shapes.
Section~\ref{sec:theory-CDFT} includes a brief overview of covariant density functional
theory the current implementation of the relativistic SCMF model is described in detail
in section~\ref{sec:numerical}, and the structure of the program is
explained in section~\ref{sec:program-structure}.
\section{Covariant density functional theory}
\label{sec:theory-CDFT}
In conventional quantum hadrodynamics
(QHD)~\cite{Serot1986_ANP16-1,Serot1997_IJMPE6-181,Lalazissis2004_LNP641,Furnstahl2000_CommentsNPPA2-23}
a nucleus is described as a system of Dirac nucleons coupled to
exchange mesons through an effective Lagrangian. The isoscalar-scalar $\sigma$ meson,
the isoscalar-vector $\omega$ meson, and the isovector-vector $\rho$ meson build the
minimal set of meson fields that is necessary for a description of bulk and single-particle
nuclear properties. Of course, at the scale of low-energy nuclear structure, heavy-meson
exchange is just a convenient representation of the effective nuclear interaction.
The behavior of the nucleon-nucleon (NN) interaction at long and intermediate
distances is determined by one- and two-pion exchange processes. The exchange of
heavy mesons is associated with short-distance dynamics that cannot be resolved at
low energies that characterize nuclear binding and, therefore, can be represented
by local four-point (contact) NN interactions,with low-energy (medium-dependent)
parameters adjusted to nuclear data.
The DIRHB program package includes both the meson-exchange and the
point-coupling representations of the relativistic nuclear energy density functional (NEDF).
\subsection{Meson-exchange models}
\label{subsec:theory-meson-exchange}
The meson-exchange model is defined by the Lagrangian density
\begin{equation}
{\cal L} = {\cal L}_N + {\cal L}_m + {\cal L}_{int}.
\end{equation}
${\cal L}_N$ denotes the Lagrangian of the free nucleon
\begin{equation}
\label{Eq:Lfree}
{\cal L}_N = \bar{\psi}(i\gamma_\mu\partial^\mu-m)\psi,
\end{equation}
where $m$ is the bare nucleon mass and $\psi$ denotes the Dirac spinor.
${\cal L}_m$ is the Lagrangian for the free meson fields and electromagnetic field
\begin{align}
{\cal L}_m &= \frac{1}{2}\partial_\mu\sigma\partial^\mu\sigma -\frac{1}{2}m^2_\sigma\sigma^2%
-\frac{1}{2}\Omega_{\mu\nu}\Omega^{\mu\nu} +\frac{1}{2}m^2_\omega\omega_\mu\omega^\mu\nonumber \\
&-\frac{1}{4}\vec{R}_{\mu\nu}\cdot\vec{R}^{\mu\nu} +\frac{1}{2}m^2_\rho\vec{\rho}_\mu\cdot\vec{\rho}^\mu%
-\frac{1}{4}F_{\mu\nu}F^{\mu\nu},
\end{align}
with the corresponding masses $m_\sigma$, $m_\omega$, $m_\rho$, and 
$\Omega_{\mu \nu}$, $\vec{R}_{\mu \nu}$, $F_{\mu \nu}$ are field tensors
\begin{align}
\Omega_{\mu\nu} &= \partial_\mu\omega_\nu-\partial_\nu\omega_\mu, \\%
\vec{R}_{\mu\nu} &= \partial_\mu\vec{\rho}_\nu-\partial_\nu\vec{\rho}_\mu,\\%
F_{\mu\nu} &= \partial_\mu A_\nu-\partial_\nu A_\mu.%
\end{align}
Arrows denote isovectors and boldface symbols are used for vectors in ordinary space.
The minimal set of interaction terms is contained in ${\cal L}_{int}$
\begin{equation}
{\cal L}_{int} = - g_\sigma\bar{\psi}\psi\sigma%
-g_\omega\bar{\psi}\gamma^\mu\psi\omega_\mu%
-g_\rho\bar{\psi}\vec{\tau}\gamma^\mu\psi\cdot\vec{\rho}_\mu%
-e\bar{\psi}\gamma^\mu\psi A_\mu,
\end{equation}
with the couplings $g_\sigma$, $g_\omega$, $g_\rho$ and $e$.

From the Lagrangian density, one can easily obtain the Hamiltonian density
(for details see Ref.~\cite{Ring1996_PPNP37-193}), which for the static case reads
\begin{align}
\label{Eq:Ham}
{\cal H}(\boldsymbol{r}) &= \sum_i^A\psi_i^\dagger \left(\boldsymbol{\alpha}\boldsymbol{p} + \beta m  \right)\psi_i\nonumber\\
&+\frac{1}{2}\left[(\boldsymbol{\nabla} \sigma)^2 + m_\sigma^2\sigma^2  \right]%
-\frac{1}{2}\left[(\boldsymbol{\nabla}\omega)^2 + m_\omega^2\omega^2  \right]\nonumber\\
&-\frac{1}{2}\left[(\boldsymbol{\nabla}\rho)^2 + m_\rho^2\rho^2  \right]%
-\frac{1}{2}(\boldsymbol{\nabla} A)^2\nonumber\\
&+\left[ g_\sigma \rho_s \sigma + g_\omega j_\mu\omega^\mu+g_\rho \vec{j}_\mu\cdot\vec{\rho}^\mu
+ej_{p\mu}A^\mu  \right].
\end{align}
We have also introduced the isoscalar-scalar density, the isoscalar-vector current,
the isovector-vector current and the electromagnetic current
\begin{align}
\rho_s(\boldsymbol{r}) &= \sum_{i=1}^A{\bar{\psi}_i(\boldsymbol{r})\psi_i(\boldsymbol{r})},\\
j_\mu(\boldsymbol{r}) &= \sum_{i=1}^A{\bar{\psi}_i(\boldsymbol{r})\gamma_\mu\psi_i(\boldsymbol{r})}, \\
\vec{j}_\mu(\boldsymbol{r}) &= \sum_{i=1}^A{\bar{\psi}_i(\boldsymbol{r})\vec{\tau}\gamma_\mu\psi_i(\boldsymbol{r})}, \\
j_{p\mu}(\boldsymbol{r}) &= \sum_{i=1}^Z{\psi^\dag_i(\boldsymbol{r})\gamma_\mu\psi_i(\boldsymbol{r})},
\end{align}
where the summation is performed only over occupied orbits in the Fermi sea of positive energy
states, i.e. the {\it no-sea} approximation is used. Although the contributions of the Dirac-sea
are not small~\cite{Chin1977_APNY108-301,Perry1986_PLB182-269,ZHU-ZY1991_PLB254-325},
they are in fact taken into account by adjusting the model parameters to the experimental 
data~\cite{Furnstahl1997_NPA618-446}.
By integrating the Hamiltonian density (\ref{Eq:Ham}) over the $r$-space we obtain
the total energy which depends on the Dirac spinors and the meson fields
\begin{equation}
\label{Eq:EDF-DD}
E_{RMF}[\psi,\bar{\psi},\sigma,\omega^\mu,\vec{\rho}^\mu,A^\mu] = %
\int d^3r\,{\cal H}(\boldsymbol{r}).%
\end{equation}

Already in the earliest applications of the relativistic mean-field (RMF) framework it was realized, 
however, that this simple
model with interaction terms only linear in the meson fields does not provide a quantitative description
of complex nuclear system~\cite{Boguta1977_NPA292-413,Pannert1987_PRL59-2420}.
Therefore, an effective density dependence was introduced
by replacing the quadratic $\sigma$-potential with a quartic one~\cite{Boguta1977_NPA292-413}.
This model has been used successfully in a number of 
studies~\cite{Reinhard1986_ZPA323-13,Gambhir1990_APNY198-132,Lalazissis1997_PRC55-540,Todd-Rutel2005_PRL95-122501}, and the corresponding computer code for
axial deformed systems has been published in Ref.~\cite{Ring1997_CPC105-77}.
Of course, implementation of the covariant density functional with non-linear meson couplings
has no direct physical meaning. Therefore, is seems more natural to follow an idea of Brockmann and
Toki~\cite{Brockmann1992_PRL68-3804} and use density-dependent couplings. $g_\sigma$,
$g_\omega$ and $g_\rho$ are assumed to be vertex functions of Lorentz-scalar bilinear forms
of the nucleon operators. In most applications the meson-nucleon couplings are functions of the
vector density $\rho_v =  \sqrt{j_\mu j^\mu}$, with the nucleon four-current 
$j^\mu = \bar{\psi}\gamma^\mu \psi = \hat{\rho}u^\mu$. $u^\mu$ is the four-velocity, defined as
$(1-\mathbf{v}^2)^{-1/2}(1,\mathbf{v})$. In the rest-frame of homogeneous nuclear matter: $\mathbf{v}=0$.
Brockmann and Toki derived the density dependence of the couplings in an ab-initio
calculation from the relativistic Brueckner-Hartree-Fock calculation in
the infinite nuclear matter. Since there are no free parameters in this model,  the results of such a calculation
are not very precise. Therefore, a phenomenological ansatz is introduced for the density-dependence of
the couplings with parameters adjusted to the experimental data in finite nuclei~\cite{Typel1999_NPA656-331,Hofmann2001_PRC64-034314,Niksic2002_PRC66-024306,Lalazissis2005_PRC71-024312}.

The single-nucleon Dirac equation is derived by variation of the energy density functional (\ref{Eq:EDF-DD}) with respect to $\bar{\psi}$
\begin{equation}
\label{Eq:var1}
\hat{h}_D \psi_i = \epsilon_i \psi_i,
\end{equation}
with the Dirac Hamiltonian
\begin{equation}
\label{Eq:Dirac0}
\hat{h}_D = \boldsymbol{\alpha}(\boldsymbol{p}-\boldsymbol{\Sigma}) + \Sigma_0 + \beta (m+\Sigma_s).
\end{equation}
The nucleon self-energies $\Sigma$ are defined by the following expressions
\begin{align}
\Sigma_s(\boldsymbol{r}) &= g_\sigma \sigma(\boldsymbol{r}), \nonumber \\
\Sigma_\mu(\boldsymbol{r}) &= g_\omega\omega_\mu(\boldsymbol{r}) + g_\rho \vec{\tau}\cdot\vec{\rho}_\mu(\boldsymbol{r})
+ eA_\mu(\boldsymbol{r}) + \Sigma_\mu^R(\boldsymbol{r}).
\end{align}
The density dependence of the vertex functions $g_\sigma$, $g_\omega$, and $g_\rho$
produces the {\it rearrangement} contribution to the vector self-energy
\begin{equation}
\label{Eq:rearr}
\Sigma_\mu^R = \frac{j_\mu}{\rho_v}\left(
\frac{\partial g_\sigma}{\partial \rho_v} \rho_s \sigma%
+\frac{\partial g_\omega}{\partial \rho_v}j_\nu\omega^\nu%
+\frac{\partial g_\rho}{\partial \rho_v} \vec{j}_\nu \cdot \vec{\rho}^\nu
\right) .
\end{equation}
The variation of the energy density functional (\ref{Eq:EDF-DD}) with respect to the meson
fields leads to the Helmholtz equations for the meson fields
\begin{align}
\label{Eq:Helmholtz}
\left[-\Delta + m_\sigma^2\right] \sigma~~&= -g_\sigma\rho_s,\\
\left[-\Delta + m_\omega^2\right] \omega^\mu &=~~~g_\omega j^\mu,\\
\left[-\Delta + m_\rho^2\right] \vec{\rho}^\mu &=~~~g_\rho \vec{j}^\mu,
\end{align}
and to the Poisson equation for the electromagnetic field
\begin{equation}
\label{Eq:Laplace}
-\Delta A^\mu = ej^\mu_p.
\end{equation}
Because of charge conservation, only the 3rd component of the isovector $\rho$-meson
contributes. In the ground-state solution for an even-even nucleus there are no currents
(time-reversal invariance), and the corresponding spatial components of the meson-fields
vanish. For simplicity, the surviving components of the meson fields are denoted by
$\rho$ and $\omega$.  The Dirac equation takes a
simple form that includes only the vector potential
$V(\boldsymbol{r})$ and the effective mass $M^*(\boldsymbol{r})=m + g_\sigma \sigma$,
\begin{equation}
\label{Eq:Dirac1}
\{-i\boldsymbol{\alpha}\boldsymbol{\nabla} + \beta M^*(\boldsymbol{r}) + V(\boldsymbol{r}) \}
\psi_i(\boldsymbol{r}) = \epsilon_i \psi_i(\boldsymbol{r}).
\end{equation}
The vector potential reads
\begin{equation}
\label{Eq:vector-potential}
V(\boldsymbol{r}) = g_\omega \omega + g_\rho \tau_3 \rho + eA_0
 +\Sigma_0^R,
\end{equation}
and the rearrangement contribution (\ref{Eq:rearr}) is reduced to 
\begin{equation}
\Sigma_0^R = \frac{\partial g_\sigma}{\partial \rho_v}\rho_s\sigma
+\frac{\partial g_\omega}{\partial \rho_v}\rho_v \omega
+\frac{\partial g_\omega}{\partial \rho_v}\rho_{tv} \rho.
\end{equation}
$\rho_{tv}$ denotes the isovector density, i.e. the difference between the proton and the neutron
vector density.
The density dependence of the meson-nucleon couplings is parameterized in a phenomenological
way \cite{Typel1999_NPA656-331,Hofmann2001_PRC64-034314,Niksic2002_PRC66-024306}.
The coupling of the $\sigma$-meson and $\omega$-meson to the nucleon field reads
\begin{equation}
g_i(\rho) = g_i(\rho_{sat})f_i(x) \quad \textrm{for} \; i=\sigma,\omega,
\end{equation}
where
\begin{equation}
\label{Eq:f-coupl}
f_i(x) = a_i \frac{1+b_i(x+d_i)^2}{1+c_i(x+d_i)^2},
\end{equation}
is a function of $x=\rho/\rho_{sat}$, and $\rho_{sat}$ denotes the baryon density at saturation in
symmetric nuclear matter. The eight real parameters in Eq. (\ref{Eq:f-coupl}) are not independent.
The five constraints
\begin{equation}
\label{Eq:constraints-f}
f_i(1)=1,\quad f_\sigma^{\prime \prime}(1)=f_\omega^{\prime \prime}(1),  \quad f_i^{\prime \prime}(0)=0,
\end{equation}
reduce the number of independent parameters to three. Three additional parameters
in the isoscalar channel are: $g_\sigma(\rho_{sat})$, $g_\omega(\rho_{sat})$, and $m_\sigma$ -- the
mass of the phenomenological $\sigma$-meson. 
For the $\rho$-meson coupling the functional form of the density
dependence is suggested by a Dirac-Brueckner calculation of asymmetric nuclear
matter \cite{DeJong1998_PRC57-3099}
\begin{equation}
g_\rho(\rho)=g_\rho(\rho_{sat}) e^{-a_\rho(x-1)}.
\end{equation}
The isovector channel is parameterized by $g_\rho(\rho_{sat})$ and $a_\rho$. For the masses of the $\omega$
and $\rho$ mesons the free values are used: $m_\omega=783$ MeV and $m_\rho=763$ MeV.
The eight independent parameters (seven coupling parameters and the mass of the $\sigma$-meson) are
adjusted to reproduce empirical properties of symmetric and asymmetric nuclear matter, binding energies,
charge radii, and neutron radii of spherical nuclei. The DIRHB program package includes
the very successful density-dependent meson-exchange relativistic energy functional
DD-ME2~\cite{Lalazissis2005_PRC71-024312}.
\subsection{Point-coupling models}
\label{subsec:theory-point-coupling}
Point-coupling models represent an alternative formulation of the self-consistent relativistic
mean-field framework \cite{Manakos1989_ZPA330-223,Rusnak1997_NPA627-495,Buervenich2002_PRC65-44308,Niksic2008_PRC78-034318,ZHAO-PW2010_PRC82-054319}. 
In complete analogy to the meson-exchange phenomenology, in
which the isoscalar-scalar $\sigma$ meson, the isoscalar-vector $\omega$ meson, and the
isovector-vector $\rho$ meson build the minimal set of meson fields that is necessary for
a quantitative description of nuclei, an effective Lagrangian that includes the isoscalar-scalar,
isoscalar-vector and isovector-vector four-fermion interaction reads
\begin{align}
\mathcal{L} &= \bar{\psi}(i\gamma \cdot \partial - m)\psi \nonumber \\
 &-\frac{1}{2}\alpha_S(\rho) (\bar{\psi}\psi)(\bar{\psi}\psi) 
 -\frac{1}{2}\alpha_V(\rho) (\bar{\psi}\gamma^\mu \psi)(\bar{\psi}\gamma_\mu \psi)
 -\frac{1}{2}\alpha_{TV}(\rho) (\bar{\psi}\vec{\tau}\gamma^\mu \psi)(\bar{\psi}\vec{\tau}\gamma_\mu \psi)
 \nonumber \\
&- \frac{1}{2}\delta_S (\partial_\nu \bar{\psi} \psi)(\partial^\nu \bar{\psi}\mu \psi)
-e\bar{\psi}\gamma \cdot A \frac{1-\tau_3}{2}\psi.
\label{eq:PC-lag}
\end{align}
In addition to the free-nucleon Lagrangian and the point-coupling interaction terms, the model includes
the coupling of protons to the electromagnetic field. The derivative term in Eq.~(\ref{eq:PC-lag}) accounts
for leading effects of finite-range interactions that are crucial for a quantitative description of nuclear
density distribution, e.g. nuclear radii. Similar interactions can be included in each space-isospace
channel, but in practice data only constrain a single derivative term, for instance 
$\delta_S(\partial_\nu \bar{\psi}\psi)(\partial^\nu \bar{\psi}\psi)$. The inclusion of an adjustable derivative
term only in the isoscalar-scalar channel is consistent with conventional meson-exchange RMF
models, in which the mass of the phenomenological $\sigma$ meson is treated as a free parameter, whereas
free values are used for the masses of the $\omega$ and $\rho$ mesons.

The couplings of the interaction terms in Eq.~(\ref{eq:PC-lag}) are functionals of the nucleon 4-current
\begin{equation}
j^\mu = \bar{\psi}\gamma^\mu \psi = \hat{\rho} u^\mu,
\end{equation}
where $u^\mu$ is the 4-velocity defined as $(1-\mathbf{v}^2)^{-1/2}(1,\mathbf{v})$. In the rest-frame
of homogeneous nuclear matter: $\mathbf{v}=0$.

Following the procedure described in section~\ref{subsec:theory-meson-exchange} we can
derive the Hamiltonian density ${\cal H}(\boldsymbol{r})$ and the EDF for the point-coupling model
\begin{align}
E_{RMF}[\psi,\bar{\psi},A_\mu] &= \int{d^3r\,{\cal H}(\boldsymbol{r})}\nonumber \\%
&= \sum_{i=1}^A{\int{ d^3r \psi_i^\dagger\left(\boldsymbol{\alpha}\boldsymbol{p} + \beta m\right)\psi_i }}%
- \frac{1}{2}(\boldsymbol{\nabla} A)^2 \nonumber + \frac{1}{2}e \int{d^3r j_p^\mu A_\mu}  \\
&+\frac{1}{2}\int{ d^3r \left[ \alpha_S \rho_s^2 + \alpha_V j_\mu j^\mu
+  \alpha_{TV}\vec{j}_\mu\cdot\vec{j}^\mu  +  \delta_S \rho_s \Box \rho_s \right]}.
\label{Eq:EDF-PC}
\end{align}
The variation of the EDF Eq.~(\ref{Eq:EDF-PC}) with respect to the Dirac spinors $\bar{\psi}$ 
leads to the Dirac equation
\begin{equation}
\label{Eq:Dirac2}
\{-i\boldsymbol{\alpha}\boldsymbol{\nabla} + \beta M^*(\boldsymbol{r}) + V(\boldsymbol{r}) \}
\psi_i(\boldsymbol{r}) = \epsilon_i \psi_i(\boldsymbol{r}),
\end{equation}
where we assume that the time-reversal invariance is valid.
The Dirac effective mass is defined
\begin{equation}
M^* = m + \Sigma_s = m + \alpha_S\rho_s + \delta_S \triangle \rho_s,
\end{equation}
while the vector potential reads
\begin{equation}
V(\boldsymbol{r}) = \alpha_V\rho_v + \alpha_{TV}\tau_3 \rho_{tv} + eA_0
 +\Sigma_0^R.
\end{equation}
$\Sigma_0^R$ denotes the {\it rearrangement} contribution, arising from the variation of the
couplings $\alpha_S$, $\alpha_V$, and $\alpha_{TV}$ with respect to the nucleon fields in the
density operator $\hat{\rho}$
\begin{equation}
\Sigma_0^R = \frac{\partial \alpha_S}{\partial \rho_v}\rho_s^2%
+\frac{\partial \alpha_V}{\partial \rho_v}\rho_v^2%
+\frac{\partial \alpha_{TV}}{\partial \rho_v}\rho_{tv}^2.
\end{equation}

The DIRHB program package includes recently developed density-dependent point-coupling
interaction DD-PC1. Here we only give brief description of the procedure used to adjust
the model parameters, whereas further details can be found in Ref.~\cite{Niksic2008_PRC78-034318}.
Guided by the Hartree-Fock scalar and vector self-energies of the Idaho next-to-next-to-next-to
leading order (N$^{3}$LO) potential~\cite{Entem.03},
we have chosen the following practical ansatz for the functional form of the couplings 
\begin{equation}
\label{eq:alpha_i}
\alpha_i (\rho) = a_i + (b_i + c_i x)e^{-d_i x},\quad (i=S,V,TV),
\end{equation}
where $x=\rho/\rho_{sat}$, and $\rho_{sat}$ denotes the nucleon density at saturation 
in symmetric nuclear matter. Although we use different formulas for the density dependence 
in the meson-exchange and point-coupling models, the scalar and vector self-energies
are similar in both models. We notice that the form Eq.~(\ref{eq:alpha_i}) is somewhat
more convenient to use when adjusting the model parameters.

In the isovector channel the corresponding Hartree-Fock {\it tree-level} nucleon self-energies,
obtained by directly mapping microscopic nucleon-nucleon potentials on a relativistic operator
basis, were not available. Therefore, the functional form of the coupling $\alpha_{TV}(\rho)$
was determined from the results of Dirac-Brueckner calculations of asymmetric nuclear 
matter~\cite{DeJong1998_PRC57-3099}, as was done in the case of the finite-range meson-exchange
interactions described in section~\ref{subsec:theory-meson-exchange}. Accordingly, for the
isovector channel we set two parameters to zero - $a_{TV}=0$ and $c_{TV}=0$ - and adjust
$b_{TV}$ and $d_{TV}$ to empirical properties of asymmetric nuclear matter and to 
nuclear masses, together with the parameters of the isoscalar channel. 

In order to reduce the number of free parameters, we have set the value $c_V=0$.
The model parameters ($a_S$, $b_S$, $c_S$, $d_S$, $a_V$, $b_V$, $d_V$
$b_{TV}$, $d_{TV}$ and $\delta_S$)  were adjusted simultaneously
to infinite and semi-infinite nuclear matter properties and to the binding energies of 64 axially symmetric
deformed nuclei in the mass regions $A\approx 150-180$ and $A\approx 230-250$ .

The resulting energy functional DD-PC1, implemented in the DIRHB program package,  has been
further tested in calculations of binding energies, charge radii,
deformation parameters, neutron skin thickness, and excitation
energies of giant monopole and dipole resonances~\cite{Niksic2008_PRC78-034318}.
\subsection{Covariant density functional theory with pairing}
\label{subsec:theory-pairing}
Relativistic energy density functionals have successfully been employed in
studies of properties of ground and excited states in spherical and deformed
nuclei. For a quantitative analysis of open-shell nuclei it is necessary to
consider also pairing correlations. Pairing has often been taken into account
in a very phenomenological way in the Bardeen-Cooper-Schrieffer (BCS) model with the monopole pairing
force, adjusted to the experimental odd-even mass differences. In many cases,
however, this approach presents only a poor approximation. The physics of
weakly-bound nuclei, in particular, necessitates a unified and self-consistent
treatment of mean-field and pairing correlations. This has led to the
formulation and development of the relativistic Hartree-Bogoliubov (RHB)
model~\cite{Kucharek1991_ZPA339-23,Ring1996_PPNP37-193},
which represents a relativistic extension of the
conventional Hartree-Fock-Bogoliubov framework. The RHB model provides a
unified description of particle-hole $(ph)$ and particle-particle $(pp)$
correlations on a mean-field level by using two average potentials: the
self-consistent mean field that encloses all the long range \textit{ph}
correlations, and a pairing field $\hat{\Delta}$ which sums up the
\textit{pp}-correlations. The ground state of a nucleus is described by a
generalized Slater determinant $|\Phi\rangle$ that represents the vacuum with
respect to independent quasiparticles. The quasiparticle operators are defined
by the unitary Bogoliubov transformation of the single-nucleon creation and
annihilation operators:
\begin{equation}
\alpha_{k}^{+}=\sum\limits_{n}U_{nk}^{{}}c_{n}^{+}+V_{nk}^{{}}c_{n}^{{}}\;,
\end{equation}
where the index $n$ refers to the original basis, e.g. an oscillator basis, or
the coordinates ($\boldsymbol{r},s,t$) in space, spin and isospin. In
addition, for the relativistic case the index $p=f,g$  will denote
the large and small components of the corresponding Dirac spinor.
$U$ and $V$ are the Hartree-Bogoliubov wave functions determined by the
variational principle. In the presence of pairing the single-particle density matrix
is generalized to two densities~\cite{Ring1980}:
the normal density $\hat{\rho}$ and the pairing tensor $\hat{\kappa}$
\begin{equation}
\label{Eq:densities}
\hat{\rho}^{}_{nn^\prime}= \langle\Phi|c^\dag_{n^\prime}c^{}_{n^{}}|\Phi\rangle,~~~~~~~~~~~~~~
\hat{\kappa}^{}_{nn^\prime}= \langle\Phi|c^{}_{n^\prime}c^{}_{n^{}}|\Phi\rangle.
\end{equation}
The RHB energy density functional thus depends on both densities
\begin{equation}
\label{Eq:EDFP}
E_{RHB}[\hat{\rho},\hat{\kappa}]=E_{RMF}[\hat{\rho}]+E_{pair}[\hat{\kappa}] \;,
\end{equation}
where $E_{RMF}[\hat{\rho}]$ is the RMF-functional defined by Eqs.~(\ref{Eq:EDF-DD})
or (\ref{Eq:EDF-PC}), and the pairing part of the RHB functional reads
\begin{equation}
\label{Eq:pairing-energy}
E_{pair}[\hat{\kappa}] = \frac{1}{4}\sum_{n^{}_1n^\prime_1}\sum_{n^{}_2n^\prime_2}
\kappa^\ast_{n^{}_1n^\prime_1} \langle n^{}_1n^\prime_1|V^{pp}|n^{}_2n^\prime_2\rangle\kappa^{}_{n^{}_2n^\prime_2}.%
\end{equation}
$\langle n^{}_1n^\prime_1|V^{pp}|n^{}_2n^\prime_2\rangle$ are the matrix elements of the two-body pairing interaction.
The RHB-coefficients $U$ and $V$ are obtained by the variational principle, which yields the RHB equations:
\begin{equation}
\label{Eq:RHB}\left(
\begin{array}
[c]{cc}%
\hat{h}_{D}-m-\lambda & \hat{\Delta}\\
-\hat{\Delta}^{*} & -\hat{h}_{D}^{*}+m+\lambda
\end{array}
\right)  \left(
\begin{array}
[c]{c}%
U_{k}\\
V_{k}
\end{array}
\right)  = E_{k} \left(
\begin{array}
[c]{c}%
U_{k}\\
V_{k}
\end{array}
\right)  \;.
\end{equation}
In the relativistic case the self-consistent mean-field corresponds to the
single-nucleon Dirac Hamiltonian $\hat{h}_{D}$ of Eqs.~(\ref{Eq:Dirac1}) or (\ref{Eq:Dirac2}).
$m$ is the nucleon mass, and the chemical potential
$\lambda$ is determined by the particle number subsidiary condition such
that the expectation value of the particle number operator in the ground state
equals the number of nucleons. The pairing field $\Delta$ reads
\begin{equation}
\Delta_{n^{}_1n^\prime_1} = \frac{1}{2}\sum_{n^{}_2n^\prime_2} \langle n^{}_1n^\prime_1|V^{pp}|n^{}_2n^\prime_2\rangle
\kappa^{}_{n^{}_2n^\prime_2}.
\end{equation}
The column vectors in the eigenvalue problem Eq.~(\ref{Eq:RHB}) denote the quasiparticle
wave functions, and $E_{k}$ are the quasiparticle energies. The dimension of the RHB matrix
equation is two times the dimension of the corresponding Dirac matrix equation. Therefore, for
each eigenvector $(U_{k} ,V_{k} )$ with positive quasiparticle energy $E_{k} >
0$, there exists an eigenvector $(V_{k}^{*},U_{k}^{*})$ with negative quasiparticle
energy $-E_{k}$. Since the baryon quasiparticle operators satisfy fermion
commutation relations, the levels $E_{k}$ and $-E_{k}$ cannot be occupied
simultaneously, that is, one chooses either the positive or the negative eigenvalue
and the corresponding eigenvector~\cite{Ring1980}. In the Hartree-Fock case, the choice of positive
or negative value of the quasiparticle energy means that the level is either occupied or empty.
For the non-relativistic HFB the ground state represents the minimum of the
energy surface and, to form a vacuum with respect to all quasiparticles,
one chooses only the positive quasiparticle energies
\begin{equation}
\alpha_k | \Phi \rangle =0 \quad \textnormal{for} \quad E_k>0 \qquad \textnormal{or}\qquad
|\Phi \rangle = \prod_{E_k>0}{\alpha_k |-\rangle}.
\end{equation}
$|\Phi\rangle$ denotes the quasiparticle vacuum, whereas $|-\rangle$ is the bare vacuum.
$|\Phi\rangle$ corresponds to the occupation of states with lowest energy, because
all the quasiparticle excitations have a positive excitation energy.
The single-particle density and the pairing tensor Eq.~(\ref{Eq:densities}) that correspond
to this state can be expressed in terms of the quasi-particle wave functions:
\begin{align}
\label{Eq:pairing-tensor}
\rho^{}_{nn^\prime} &= \sum_{E_k>0}V^{*}_{nk}V_{n^\prime k},\\
\kappa^{}_{nn^\prime} &= \sum_{E_k>0}V^{*}_{nk}U_{n^\prime k}.
\end{align}
In the relativistic case one finds solutions in the Dirac sea (usually called negative energy
solutions and denoted by the index $a$), and solutions above the Dirac sea (usually
called positive energy solutions and denoted by the index $p$). In the Dirac equation
without pairing they can easily be distinguished by the sign of the corresponding eigenvalues.
For the RHB equations~(\ref{Eq:RHB}) this is no longer the case but, because
of the large gap between the Dirac and the Fermi sea ($\approx 1200$ MeV),
one can easily distinguish the levels in the Dirac sea $|E_a|>1200$ MeV, from those above
the Dirac sea (note that for the Hamiltonian $\hat{h}-m$ the
positive energy continuum corresponds to zero energy and the negative
energy continuum corresponds to $-2m$). The {\it no-sea approximation} means
that we have to choose solutions with positive quasiparticle energies $E_p>0$
for the states above the Dirac sea, and solutions with negative quasiparticle
energies $E_a<0$ for all the levels in the Dirac sea. Therefore, the RHB
ground state is given by
\begin{equation}
|\Phi \rangle = \prod_{E_p > 0}{\alpha_p \prod_{E_a < 0}{
\alpha_a |-\rangle}   }.
\end{equation}
Finally, we find for this wave function the following expressions for the single-particle
density and the pairing tensor:
\begin{align}
\label{Eq:rho-no-sea}
\rho^{}_{nn^\prime} &= \sum_{E_p>0}{V^*_{np}V_{n^\prime p} }
   +\sum_{E_a<0}{V^*_{na}V_{n^\prime a} } ,\\
\label{Eq:kappa-no-sea}
\kappa^{}_{nn^\prime} &=    \sum_{E_p>0}{V^*_{np}U_{n^\prime p} }
   +\sum_{E_a<0}{V^*_{na}U_{n^\prime a} }.
\end{align}
It should be noted that the coefficients $V_{na}$ are negligible because of the
large Dirac gap and, therefore, the second term in Eqs.~(\ref{Eq:rho-no-sea})
and (\ref{Eq:kappa-no-sea}) is often neglected. In this case, however,
the pairing tensor $\hat{\kappa}$ is only approximately antisymmetric.

Pairing correlations in nuclei are restricted to an energy window of a few MeV
around the Fermi level, and their scale is well separated from the scale of
binding energies, that are in the range from several hundred to thousand MeV.
There is no empirical evidence for any relativistic effect in the nuclear
pairing field $\hat{\Delta}$ and, therefore, a hybrid RHB model~\cite{GonzalesLlarena1996_PLB379-13} with a
non-relativistic pairing interaction can be formulated. For a general two-body
interaction, the matrix elements of the relativistic pairing field read
\begin{equation}
\hat{\Delta}_{n_1 p_1, n^\prime_1 p^\prime_1} = {\frac{1}{2}}\sum\limits_{n_2 p_2, n^\prime_2 p^\prime_2}
\langle n_1 p_1, n^\prime_1 p^\prime_1|V^{pp}|n_2 p_2, n^\prime_2 p^\prime_2\rangle \kappa_{n_2 p_2,n^\prime_2 p^\prime_2}.
\end{equation}
The indices ($p_{1},p_{2},p_{3},p_{4} \equiv f, g$) refer to the large
and small components of the quasiparticle Dirac spinors:
\begin{equation}
U_{k}\ =\ \left(
\begin{array}
[c]{c}%
f^{(U)}_{k}\\
ig^{(U)}_{k}
\end{array}
\right)
\quad\quad\quad (V)_k\ =\ \left(
\begin{array}
[c]{c}%
f^{(V)}_{k}\\
ig^{(V)}_{k}
\end{array}
\right)  \; . \label{Eq:UV}%
\end{equation}
In practical applications of the RHB model only
the large components of the spinors $U_{k}$ and $V_{k}$ are used to build
the non-relativistic pairing tensor
$\hat{\kappa}$ in Eq.~(\ref{Eq:pairing-tensor}). The resulting pairing field
reads
\begin{equation}
\label{Eq:pairing-matrix}
\hat{\Delta}_{n^{}_{1} f, n^\prime_{1} f} = {\frac{1}{2}}\sum\limits_{n^{}_2n^\prime_2}
\langle n^{}_{1} f, n^\prime_{1} f |V^{pp}|n^{}_{2} f, n^\prime_{2} f\rangle_{a}~ \kappa_{n^{}_{2} f, n^\prime_{2} f} .
\end{equation}
The other components: $\hat{\Delta}_{fg}$, $\hat{\Delta}_{gf}$, and
$\hat{\Delta}_{gg}$ can be safely omitted~\cite{Serra2002_PRC65-064324}.

For reasons of simplicity in many RHB calculations, in particular for those which
serve as a basis for investigations beyond mean field~\cite{Niksic2011_PPNP66-519} 
a zero-range pairing force has been chosen. However, this force shares with
the monopole force with constant $G$ the problem of an ultraviolet divergence
and requires the use of a pairing window. Finite range forces avoid this. 
Therefore the pairing part of the Gogny force D1S~\cite{Berger1991_CPC61-365} 
has been used with great success for many conventional RHB calculations for 
nuclear ground state properties~\cite{GonzalesLlarena1996_PLB379-13}, 
for the study of rotational bands in the rotating frame~\cite{Afanasjev1999_PRC60-051303},
and for investigations of giant resonances in the framework of the relativistic quasiparticle
random phase approximation (RQRPA)~\cite{Vretenar2005_PR409-101}. 
Since the calculations involving the finite-range Gogny force in the pairing channel
require considerable computational effort, a separable form of the Gogny
force has been introduced for RHB calculations in spherical and deformed nuclei~\cite{Tian2009_PLB676-44,TIAN2009_PRC79-064301,Tian2009_PRC80-024313,Niksic2010_PRC81-054318}. The force is separable in momentum space, and is completely determined by two parameters that are 
adjusted to reproduce the pairing gap of the Gogny force in symmetric nuclear matter. 
The gap equation in the $^{1}$S$_{0}$
channel reads
\begin{equation}
\Delta(k)=-\int_{0}^{\infty}{\frac{{k^{\prime2}dk^{\prime}}}{{2\pi^{2}}}%
}\left\langle k\right\vert V^{^{1}S_{0}}\left\vert k^{\prime}\right\rangle
{\frac{{\Delta(k^{\prime})}}{{2E(k^{\prime})}}}\;,
\end{equation}
and the pairing force is separable in momentum space
\begin{equation}
\left\langle k\right\vert V^{^{1}S_{0}}\left\vert k^{\prime}\right\rangle
=-Gp(k)p(k^{\prime})\;.%
\label{Eq:sep_pair}%
\end{equation}
By assuming a simple Gaussian ansatz $p(k)=e^{-a^{2}k^{2}}$, the two
parameters $G$ and $a$ have been adjusted to reproduce the density dependence
of the gap at the Fermi surface, calculated with a Gogny force. For the D1S
parameterization~\cite{Berger1991_CPC61-365} of the Gogny force the following
values were determined: $G=728\;\mathrm{MeV fm}%
^{3}$ and $a=0.644\;\mathrm{fm}$. When the pairing force Eq.~(\ref{Eq:sep_pair})
is transformed from momentum to coordinate space, it takes the form:
\begin{equation}
V^{pp}(\boldsymbol{r}_{1},\boldsymbol{r}_{2},
{\mbox{\boldmath $r$}}_{1}^{\prime},{\mbox{\boldmath $r$}}_{2}^{\prime})%
=-G\delta\left({\mbox{\boldmath $R$}}-{\mbox{\boldmath $R$}}^{\prime}\right)
P({\mbox{\boldmath $r$}})P({\mbox{\boldmath $r$}}^{\prime}),%
\label{Eq:Vpp}%
\end{equation}
where
${\mbox{\boldmath $R$}}=\frac{1}{\sqrt{2}}\left( {\mbox{\boldmath $r$}}_{1}+{\mbox{\boldmath $r$}}_{2}\right)$
and ${\mbox{\boldmath $r$}}=\frac{1}{\sqrt{2}}({\mbox{\boldmath $r$}}_{1}-{\mbox{\boldmath $r$}}_{2})$ denote the
center-of-mass and the relative coordinates, respectively, and $P({\mbox{\boldmath $r$}})$ is the Fourier transform of $p(k)$:
\begin{equation}
P({\mbox{\boldmath $r$}})=\frac{1}{\left(  4\pi a^{2}\right)  ^{3/2}%
}e^{-{\mbox{\boldmath $r$}}^{2}/2a^{2}}\;.%
\label{Eq:P3D}%
\end{equation}
The pairing force has a finite range and, because of the presence of the factor
$\delta\left(  {\mbox{\boldmath $R$}}-{\mbox{\boldmath
$R$}}^{\prime}\right)  $, it preserves translational invariance. Even though
$\delta\left(  {\mbox{\boldmath $R$}}-{\mbox{\boldmath
$R$}}^{\prime}\right)  $ implies that this force is not completely separable
in coordinate space, we will show in the following sections that the corresponding
antisymmetrized $pp$ matrix elements
\begin{equation}
\langle n^{}_1 {n}^{}_2  \vert V^{pp}  \vert n^\prime_1 {n}^\prime_2 \rangle_{a}%
=\langle  n^{}_1 {n}^{}_2  \vert V^{pp} \vert n^\prime_1 {n}^\prime_2 \rangle
-\langle  n^{}_1 {n}^{}_2  \vert V^{pp} \vert n^\prime_2 {n}^\prime_1 \rangle,
\end{equation}
can be represented as a sum of a finite number of separable terms in the
harmonic oscillator basis:
\begin{equation}
\langle n^{}_1 {n}^{}_2  \vert V^{pp}  \vert n^\prime_1 {n}^\prime_2 \rangle_{a}%
=\sum_N W^{N*}_{n^{}_1 {n}^{}_2} W^N_{n^\prime_1 {n}^\prime_2} \;.
\end{equation}
In this case the pairing field $\Delta$ takes the form
\begin{equation}
\Delta_{n^{}_1 {n}^{}_2} =\sum_N P_N\,W^{N*}_{n^{}_1 {n}^{}_2}
\qquad \textnormal{with}\quad P_N =\frac{1}{2}{\rm Tr} (W^{N}\kappa),
\end{equation}
and, finally, the pairing energy in the nuclear ground state
is given by~\cite{Tian2009_PLB676-44}:
\begin{equation}
E_{\textnormal{pair}} = -G\sum_N{P_N^* P^{}_N}.
\end{equation}

It should be noticed that the procedure to adjust the EDF parameters for the 
DD-ME2 and DD-PC1 sets has been performed by treating the pairing correlations in
the BCS constant-gap approximation with empirical pairing gaps (5-point formula).
This approximation is justified because pairing correlations contribute only a very small
portion to the total binding energy. In nuclei there is a clear separation of scales between
the bulk contributions to the binding energies of the order of hundreds to more than
thousand MeV, and the pairing energy of the order of ten MeV. To take into account
pairing correlations in a calculation of the binding of nuclei close to $\beta$-stability,
such as those used to adjust the EDF parameters, it is sufficient to consider only the monopole part 
of the effective pairing interaction adjusted to experimental pairing gaps.  Of course, this
is no longer true in studies of phenomena determined by structure effects in the vicinity
of the Fermi surface, such as nuclear excitations or fission barriers, or in nuclei far from
stability, where detailed properties of the effective interaction in the pairing channel become
important. 
\section{Numerical implementation of the RHB equations}
\label{sec:numerical}
For nuclei with spherical symmetry the RHB equation in coordinate space reduces to a simple set of radial
integro-differential equations. In the case of deformed nuclei, however, the solution of integro-differential
equations in coordinate space presents a numerically intensive and time-consuming task.
For an efficient implementation of the RHB model the DIRHB package uses a method proposed by Vautherin \cite{Vautherin1973_PRC7-296}, that combines the configurational and coordinate space representations. The RHB equation is solved in the configurational space of harmonic oscillator wave functions with appropriate symmetry, whereas the densities are computed in coordinate space. The method can be applied to spherical, axially and non-axially deformed nuclei.
The RHB eigenvalue problem in configurational space reads~\cite{Gambhir1990_APNY198-132}
\begin{equation}%
\label{Eq:Dirac3}
\left( \begin{array}{cccc}
\mathcal{A}-\lambda   & \mathcal{B}    & \Delta_{ff} & 0 \\
\mathcal{B}^T   & \mathcal{C} -\lambda & 0 & 0 \\
\Delta_{ff} & 0 & -\mathcal{A}+\lambda  &- \mathcal{B}  \\
0  &  0 & -\mathcal{B}^T  & -\mathcal{C} +\lambda
\end{array} \right)
\left( \begin{array}{c} f^{(U)} \\ g^{(U)} \\ f^{(V)} \\ g^{(V)} \end{array} \right) = E
\left( \begin{array}{c} f^{(U)} \\ g^{(U)} \\ f^{(V)} \\ g^{(V)} \end{array} \right) .
\end{equation}
The diagonalization of the RHB matrix equation yields the wave functions in
configurational space. The resulting density matrix is computed in configurational
space
\begin{equation}
\left( \begin{array}{cc} \rho_{nn^\prime} &  \rho_{n \tilde{n}^\prime}
\\  \rho_{\tilde{n} n^\prime} &  \rho_{\tilde{n} \tilde{n}^\prime}
\end{array} \right)
 = \left(  \begin{array}{cc}
~~~\sum{f^{(V)*}_{n^{ }}     f^{(V)}_{n^\prime}} &i\sum{f^{(V)*}_n          g^{(V)}_{\tilde{n}^\prime}}\\%
-i\sum{g^{(V)*}_{\tilde{n}} f^{(V)}_{n^\prime}} &~~\sum{g^{(V)*}_{\tilde{n}}g^{(V)}_{\tilde{n}^\prime}}%
\end{array} \right).
\end{equation}
where $n$ and $\tilde{n}$ denote the indices of an expansion of the large and small components
of the Dirac spinor in the oscillator basis. The density matrix is then transformed to the coordinate space,
and the resulting vector and scalar densities are used to calculate the potentials.

The map of the energy surface as a function of quadrupole deformation parameters is
obtained by solving the RHB equation with constraints on the axial and triaxial
mass quadrupole moments of a given nucleus. The
method of quadratic constraints uses an unrestricted variation of the function
\begin{equation}
\langle \hat{H} \rangle + \sum_{\mu =0,2}{C_{2\mu} ( \langle \hat{Q}_{2\mu} \rangle
-q_{2\mu})^2},
\end{equation}
where $\langle \hat{H} \rangle$ is the total energy and $\langle \hat{Q}_{2\mu}\rangle$
denotes the expectation value of the mass quadrupole operators
\begin{equation}
\hat{Q}_{20} = 2z^2-x^2 - y^2 \quad \textnormal{and} \quad
\hat{Q}_{22} = x^2 - y^2 .
\end{equation}
$q_{2\mu}$ is the constrained value of the multipole moment and $C_{2\mu}$ the
corresponding stiffness constant~\cite{Ring1980}. For a self-consistent solution the
quadratic constraint adds an extra force term $\sum_{\mu=0,2}{\lambda_\mu \hat{Q}_{2\mu}}$
to the system, where
$\lambda_\mu=2C_{2\mu}(\langle \hat{Q}_{2\mu}\rangle -q_{2\mu})$. Such a term is
necessary to force the system to a point in deformation space different from the stationary
point. In general, the values of the quadrupole moments $\langle \hat{Q}_{2\mu}\rangle$
for the self-consistent solution coincide with the constrained values $q_{2\mu}$ only
at the stationary point. Moreover, the difference between the quadrupole
moment $\langle \hat{Q}_{2\mu}\rangle$ and the constrained value $q_{2\mu}$
depends on the value of the stiffness constant, that is, smaller values of $C_{2\mu}$ lead
to larger deviations of the quadrupole moment from the corresponding constrained
value. Increasing the value of the stiffness constant, however, often destroys the
convergence of the self-consistent procedure. This deficiency can be resolved by
using the augmented Lagrangian method~\cite{Staszack2010_EPJA46-85},
and this approach has been implemented in the DIRHB package. In the iterative
procedure that leads to the self-consistent solution, the intermediate solutions can be combined by
using either the linear or the Broyden mixing procedure~\cite{Baran2008_PRC78-014318}.

\subsection{The spherically symmetric case}
\label{subsection:spherical}
For systems with rotational invariance we employ the spherical coordinates
\begin{equation}
x=r\sin{\theta}\cos{\phi},\quad y=r\sin{\theta}\sin{\phi},\quad z=r\cos{\theta}.
\end{equation}
The nucleon densities and meson fields depend only on the radial coordinate $r$.
The spinor is labeled by the nucleon angular momentum $j_i$, its projection $m_i$,
parity $\pi_i$, and the isospin projection $t_i=\pm 1/2$ for neutrons and protons, respectively
\begin{equation}
\psi_i(\boldsymbol{r},s,t) = \left( \begin{array}{c}
f_i(r)\Phi_{l_ij_im_i}(\theta,\phi,s) \\ ig_i(r)\Phi_{\tilde{l}_ij_im_i}(\theta,\phi,s)
\end{array}\right) \chi_{t_i}(t) .
\end{equation}
The orbital angular momenta that correspond to the large ($l_i$) and small ($\tilde{l}_i$) spinor
components are determined by the total angular momentum $j_i$ and parity $\pi_i$
\begin{equation}
l = j\pm \frac{1}{2}, \quad \tilde{l} = j \mp \frac{1}{2} ,\quad \pi = (-)^{j\pm\frac{1}{2}},\quad \kappa=\pm(j+\frac{1}{2}).
\end{equation}
$\chi_{t_i}$ denotes the isospin wave function, and $\Phi_{ljm}$ is a two-dimensional
spinor with the angular momentum quantum numbers $ljm$
\begin{equation}
\Phi_{ljm}(\theta,\phi,s) = \left[ \chi_{1/2}(s)\otimes Y_l(\theta,\phi)\right]_{jm}
\end{equation}
The dependence on the angles is analytical and one is left with a coupled set of ordinary
differential equations in the radial variable $r$ for the large and small components
of the Dirac spinor
\begin{align}
\left(M^*(r) + V(r) \right)f_i(r) + \left( \partial_r - \frac{\kappa_i-1}{r} \right)g_i(r)
&= \epsilon_i f_i(r),\\
- \left( \partial_r + \frac{\kappa_i+1}{r} \right)f_i(r) -\left(M^*(r) - V(r)\right)g_i(r)
&=\epsilon_i g_i(r) ,
\end{align}
where  $M^*(r)$ is the Dirac mass
and the potential $V(r)$ is defined by Eq. (\ref{Eq:vector-potential}).

The large and small components of the Dirac spinors are expanded separately in
terms of the radial functions $R_{nl}(r,b_0)$ of a spherical harmonic oscillator potential
with oscillator frequency $\hbar \omega_0$ and the corresponding oscillator
length $b_0=\sqrt{\hbar/m\omega_0}$\footnote{$m$ is the bare nucleon mass}
\begin{equation}
\label{Eq:exp-sph}
f_i(r) = \sum_{n=0}^{n_{max}}{f_n^{(i)}R_{nl_i}(r,b_0)}, \quad
g_i(r) = \sum_{\tilde{n}=0}^{\tilde{n}_{max}}{g_{\tilde{n}}^{(i)}R_{\tilde{n}\tilde{l}_i}(r,b_0)} .
\end{equation}
The radial oscillator wave functions read
\begin{equation}
R_{nl}(r,b_0) = b_0^{-3/2}R_{nl}(\xi)=b_0^{-3/2}\mathcal{N}_{nl}\,\xi^l L_{n}^{l+1/2}(\xi^2)e^{-\xi^2/2},
\end{equation}
where $\xi=r/b_0$ corresponds to the radial distance in units of the oscillator
length. $n=0,1,2,\dots$ counts the number of radial nodes. The associated Laguerre polynomials $L_n^m(\xi^2)$ are defined in
Ref.~\cite{Abramowitz1970}. The normalization factor is
\begin{equation}
\mathcal{N}_{nl} = (2n!/(l+n+1/2)!)^{1/2}.
\end{equation}
The upper limits $n_{max}$ and $\tilde{n}_{max}$ in Eq. (\ref{Eq:exp-sph}) are determined by
the corresponding major shell quantum numbers $N_{max}=2n_{max}+l_{max}$ and
$\tilde{N}_{max}=2\tilde{n}_{max}+\tilde{l}_{max}$. The small components are expanded
up to $\tilde{N}_{max}=N_{max}+1$ to avoid spurious contributions to the solution of the
RHB equation \cite{Ring1997_CPC105-77}.
\subsubsection{The Dirac Hamiltonian}
In the following the generic notation for basis states reads: $|\alpha\rangle=|nljm\rangle$. We use $\alpha$ for the expansion of the upper, and $\tilde{\alpha}$ for the expansion of the lower components of the Dirac spinor. The matrix elements of the Dirac Hamiltonian Eq.~(\ref{Eq:Dirac3}) for the case of spherical symmetry read
\begin{align}
\mathcal{A}_{\alpha \alpha^\prime}
&=\int_0^\infty{d\xi\,R_{n^{}l}(x) R_{n^{\prime} l}(\xi)} \left[ M^*(b_0\xi) + V(b_0\xi) \right],  \\
\mathcal{C}_{\tilde{\alpha}\tilde{\alpha}^\prime}
&=\int_0^\infty{d\xi\,R_{n^{}\tilde{l}}(x) R_{n^{\prime} \tilde{l}}(\xi)} \left[ M^*(b_0\xi) - V(b_0\xi) \right],  \\
\mathcal{B}_{\tilde{\alpha}\alpha^\prime} &= \mathcal{N}_{n \tilde{l}} \mathcal{N}^{}_{n^\prime l}
\int_0^\infty{ d\xi e^{-\xi^2} \xi^{2l} L_n^{\tilde{l}+1/2} L_{n^\prime}^{l+1/2}
  (2n^\prime +l +1 +\kappa -\xi^2)}.
\end{align}
\subsubsection{The Coulomb interaction}
The potential for protons includes the direct Coulomb field
\begin{equation}
V_C (\boldsymbol{r}) = e^2
\int{d^3r^\prime \frac{\rho_p(\boldsymbol{r}^\prime)}{|\boldsymbol{r}-\boldsymbol{r}^\prime  | } }.
\end{equation}
The logarithmic singularity in the integrand at the point $\boldsymbol{r}=\boldsymbol{r}^\prime$
can be eliminated by using the identity~\cite{Vautherin1973_PRC7-296}:
\begin{equation}
\triangle_{\boldsymbol{r}^\prime}|\boldsymbol{r}-\boldsymbol{r}^\prime| = \frac{2}{|\boldsymbol{r}-\boldsymbol{r}^\prime|},
\end{equation}
that, together with an integration by parts, yields
\begin{equation}
V_C(\boldsymbol{r}) = \frac{e^2}{2}
\int{d^3r^\prime |\boldsymbol{r}-\boldsymbol{r}^\prime  |
\triangle_{\boldsymbol{r}^\prime} \rho_p(\boldsymbol{r}^\prime)  }.
\end{equation}
The angular part can be integrated analytically, while the
remaining radial factor has to be integrated numerically
\begin{equation}
V_C(r) = \pi e^2
\int_0^\infty{d r^\prime {r^\prime}^2 \left(3 r + \frac{{r^\prime}^2}{ r} \right)
\frac{ d^2\rho_p(r^\prime)}{d{r^\prime}^2  }  }.
\end{equation}

\subsubsection{Klein-Gordon equations}
In spherical symmetry the Helmholtz equations for the meson fields $\phi$ = $\sigma$, $\omega$, $\rho$ read
\begin{equation}
\left( -\frac{\partial^2}{\partial r^2} - \frac{2}{r}\frac{\partial}{\partial r} + m_\phi^2\right)
\phi(r) = s_\phi(r).
\end{equation}
The solution of this equation is obtained by an expansion in a complete set of basis states
\begin{equation}
\label{Eq:KG-expansion-spherical}
\phi(r) = \sum_{n=0}^{n_b}{\phi_n R_{n0}(r,b_0)  },\quad
s_\phi(r) =  \sum_{n=0}^{n_b}{s^{\phi}_n R_{n0}(r,b_0)  }\;.
\end{equation}
The maximal radial quantum number $n_b$ in the expansion is determined
by the cut-off parameter $N_B=2n_b$.
Inserting the ansatz (\ref{Eq:KG-expansion-spherical}) into the Klein-Gordon equation, one
obtains a set of inhomogeneous linear equations,
\begin{equation}
\label{Eq:KG-spherical}
\sum_{n^\prime}^{n_b}{ \mathcal{H}_{nn^\prime} \phi_{n^\prime} } = s_n^\phi ,
\end{equation}
with the matrix
\begin{align}
\mathcal{H}_{nn^\prime} &= -b_0^{-2}\delta_{nn^\prime}  \left( 2n+3/2 \right) \nonumber\\
&+b_0^{-2}\delta_{nn^\prime+1} \sqrt{(n+1)(n+3/2)}
+b_0^{-2}\delta_{n^\prime n+1} \sqrt{(n^\prime+1)(n^\prime+3/2)}.
\end{align}
The set of equations (\ref{Eq:KG-spherical}) is solved by inversion.
\subsubsection{Pairing matrix elements}
The antisymmetric matrix elements of the pairing interaction Eq.~(\ref{Eq:Vpp}) can be separated into
products of spin and coordinate-space factors
\begin{equation}
\label{Eq:Vpp-sph}
\langle \alpha \bar{\beta}| V^{pp} | \gamma \bar{\delta} \rangle_a =%
-G\langle \alpha \bar{\beta}|
\delta\left({\mbox{\boldmath $R$}}-{\mbox{\boldmath $R$}}^{\prime}\right)
P({\mbox{\boldmath $r$}})
(1-P^\sigma)/2 | \gamma \bar{\delta} \rangle_a  .
\end{equation}
For the ground-state solution one only needs the two-particle wave functions
coupled to angular momentum $J=0$. Starting from the basis
\begin{align}
| \alpha \rangle = | nljm \rangle &= R_{nl}(r,b_0) \left[ \chi_{1/2}\otimes Y_l(\hat{\boldsymbol{r}})\right]_{jm},\\
| \bar{\alpha} \rangle = |\overline{nljm} \rangle &= (-1)^{l+j-m}R_{nl}(r,b_0) \left[ \chi_{1/2}\otimes
Y_l(\hat{\boldsymbol{r}})\right]_{j-m},
\end{align}
the two-particle wave function can be transformed from the $jj$- to the $LS$ coupling scheme. The projector
$(1-P^\sigma)/2$, together with the condition $J=0$, restricts the spin and orbital quantum numbers to
$S=\lambda=0$, respectively
\begin{equation}
|\gamma \bar{\delta}\rangle^{S=0,J=0}
=(-1)^{l_\delta} \frac{\hat{j}_\delta}{\hat{s}\hat{l}_\delta}
\delta_{l_\gamma,l_\delta}\delta_{j_\gamma,j_\delta}
R_{n_\gamma l_\gamma}(r_1,b_0)R_{n_\delta l_\delta}(r_2,b_0)
 |\lambda =0\rangle |S=0\rangle ,
\end{equation}
with $\hat{j}=\sqrt{2j+1}$. The expressions
\begin{equation}
|\lambda =0\rangle =
\left[ Y_{l_\gamma}(\hat{\boldsymbol{r}}_1) \otimes Y_{l_\delta}(\hat{\boldsymbol{r}}_2) \right]_0
\quad {\rm and}\quad
|S=0\rangle = \left[ \chi_{1/2} \otimes \chi_{1/2}\right]_0
\end{equation}
denote the angular momentum and spin wave functions coupled to angular momentum $\lambda=0$ and spin $S=0$.
The spatial wave functions are expressed in laboratory coordinates, whereas the
separable pairing interaction Eq.~(\ref{Eq:Vpp-sph}) depends on the center-of-mass coordinate
and the relative coordinate of a nucleon pair. The transformation between the laboratory and
center-of-mass reference frames can be carried out by using the Talmi-Moshinsky
brackets~\cite{Talmi1952_HelvPhysActa25-185,Moshinsky1959_NP13-104,Brody1960_NP17-16}.
In particular, the definition
of Baranger~\cite{Baranger1966_NPA79-403} is used
\begin{equation}
| n_\gamma l_\gamma n_\delta l_\delta; \lambda \mu \rangle
= \sum_{NLnl}{M^{NL~nl}_{n_\gamma l_\gamma n_\delta l_\delta} |NL nl; \lambda \mu\rangle },
\end{equation}
with
$|\lambda\mu\rangle=\left[Y_{L}(\hat{\boldsymbol{R}})\otimes Y_{l}(\hat{\boldsymbol{r}})\right]_{\lambda\mu}$%
and $|NL nl\rangle = R_{NL}(R,b_0) R_{nl}(r,b_0)$.
$M_{n_\gamma l_\gamma n_\delta l_\delta}^{NL~nl}$ are the Talmi-Moshinsky brackets,
and the conservation of harmonic oscillator quanta yields the selection rule
\begin{equation}
\label{Eq:selection-TM-spherical}
2N+L+2n+l = 2n_\gamma+l_\gamma + 2n_\delta + l_\delta .
\end{equation}
Finally, the pairing matrix elements of the interaction Eq.~(\ref{Eq:Vpp-sph}) can be expressed
as a sum over the quantum numbers $N$, $L$, $N^\prime$, $L^\prime$, $n$, $l$,
$n^\prime$ and $l^\prime$. The integration over the center of the mass coordinates
$\boldsymbol{R}$ and $\boldsymbol{R}^\prime$ yields $N=N^\prime$, $L=L^\prime$. Furthermore,
the sum contains the integrals over the relative coordinates
\begin{equation}
\int{R_{nl}(r)Y_{lm}(\hat{\boldsymbol{r}})P(r)}d^3r,
\end{equation}
where $P(r)$ is given by Eq. (\ref{Eq:P3D}). This integral vanishes for $l\neq 0$.
Since the spherical harmonics $Y_{L}(\hat{\boldsymbol{R}})$ and
$Y_{l}(\hat{\boldsymbol{r}})$ are coupled to $\lambda=0$, the condition $l=0$ also
implies $L=0$. The quantum numbers $n$ and $n^\prime$ are determined by
the selection rule Eq.~(\ref{Eq:selection-TM-spherical}), that is, a single sum
of separable terms over the quantum number $N$ is obtained
\begin{equation}
V_{\alpha \bar{\beta} \gamma \bar{\delta}}^{J=0} =
-G \sum_N W^{N*}_{\alpha \bar{\beta}} W^N_{\gamma \bar{\delta}} .
\end{equation}
The single-particle matrix elements read
\begin{equation}
W^N_{\gamma \bar{\delta}} = (-1)^{l}\frac{\hat{j}}{\hat{s}\hat{l}}M_{n_\gamma l n_\delta l}^{N0n0}
\int_0^\infty{R_{n0}(r,b_0)P(r)r^2dr  },
\end{equation}
with $l\equiv l_\gamma = l_\delta$ and $j\equiv j_\gamma = j_\delta$.
The radial integral over the relative coordinate can easily be reduced to the following form
\begin{equation}
\label{Eq:sph-radial-integral}
\int_0^\infty{R_{n0}(r,b_0)P(r)r^2dr  } = \frac{1}{b_0^{3/2}}\frac{1}{2^{1/4}}\frac{\mathcal{N}_{n0}}{(4\pi \alpha^2)^{3/2}}
 \int_0^\infty{\eta^{1/2}L_n^{1/2}(\eta)
 e^{-\frac{\eta}{2}\left(1+\frac{1}{\alpha_0}\right)}d\eta },
\end{equation}
where the parameter $\alpha_0=a/b_0$ characterizes the width of the function $P(r)$ Eq.~(\ref{Eq:P3D}) in units
of the oscillator length $b_0$.
The integral in Eq.~(\ref{Eq:sph-radial-integral})
can be computed analytically by using the generating function for
the associated Laguerre polynomials~\cite{Abramowitz1970}
\begin{equation}
\frac{e^{\frac{\eta z}{z-1}}}{(1-z)^{3/2}} = \sum_{n=1}^\infty{L_n^{1/2}(\eta)z^n  }.
\end{equation}
Finally,
\begin{equation}
W_{\gamma \bar{\delta}}^N
=\frac{(-)^l}{b_0^{3/2}}\frac{\hat{j}}{\hat{s}\hat{l}} M_{n_{\gamma}l n_{\delta}l }^{N0n0}
\frac{\sqrt{(2n+1)!}}{(2\pi)^{3/4} 2^{n}n!}
\left(\frac{1}{1+\alpha_0^2}\right)^{3/2}
\left(\frac{1-\alpha_0^2}{1+\alpha_0^2}\right)^n,
\end{equation}
where the quantum number $n$ is determined by the selection rule $n=n_\gamma+n_\delta+l-N$.

\subsection{Nuclei with axially symmetric quadrupole deformation}
\label{subsection:axial}

In the case of quadrupole deformation with axialy symmetry,  the third component $J_z$ of the total
angular momentum is conserved and defines the quantum number $\Omega_i$ that
labels the nucleon spinor
\begin{equation}
\psi_i(\boldsymbol{r},s,t) = \left(
\begin{array}{c} f_i^+(r_\perp,z) e^{i\Lambda_- \phi} \\ f_i^-(r_\perp,z) e^{i\Lambda_+ \phi}\\
ig_i^+(r_\perp,z) e^{i\Lambda_- \phi} \\ ig_i^-(r_\perp,z) e^{i\Lambda_+ \phi}
\end{array}\right) \chi_{t_i}(t)\;,
\end{equation}
where $\Lambda_{\pm}=\Omega_i\pm 1/2$, and $\{r_\perp,z,\phi\}$ are the standard cylindrical coordinates
\begin{equation}
x=r_\perp\cos{\phi}, \quad y=r_\perp\sin{\phi}, \quad z=z.
\end{equation}
Furthermore, the assumption is that parity and the third component of the isospin are conserved.
The nucleon wave functions are expanded in a basis of
eigenfunctions of a single-particle Hamiltonian for an axially deformed
harmonic oscillator potential:
\begin{equation}
V_{osc}(z,r_\perp)=\frac{1}{2}m\omega_z^2z^2 + \frac{1}{2}m\omega_\perp^2r_\perp^2.
\end{equation}
Imposing volume conservation, the two oscillator frequencies $\hbar \omega_\perp$ and
$\hbar \omega_z$ can be expressed in terms of a deformation parameter $\beta_0$,
\begin{equation}
\hbar \omega_z = \hbar \omega_0 e^{-\sqrt{\frac{5}{4\pi}}\beta_0},\quad
\hbar \omega_\perp = \hbar \omega_0 e^{\frac{1}{2}\sqrt{\frac{5}{4\pi}}\beta_0} .
\end{equation}
The corresponding oscillator length parameters are
\begin{equation}
b_z=\sqrt{\frac{\hbar}{m\omega_z}}, \quad b_\perp=\sqrt{\frac{\hbar}{m\omega_\perp}} .
\end{equation}
$b_\perp^2b_z=b_0^3$ because of volume conservation. The basis is now
determined by the two constants $\hbar \omega_0$ and $\beta_0$. The eigenfunctions
of the deformed harmonic oscillator potential are labeled by the set of quantum numbers
\begin{equation}
|\alpha \rangle = | n_z n_r \Lambda m_s\rangle,
\end{equation}
where $n_z$ and $n_r$ are the number of nodes in the $z$- and $r_\perp$-directions, respectively.
$\Lambda$ and $m_s$ are projections of the orbital angular momentum and spin on the
intrinsic $z$-axis, respectively. Making use of the dimensionless variables
\begin{equation}
\xi = z/b_z,\quad \eta=r^2_\perp/b^2_\perp ,
\end{equation}
the harmonic oscillator eigenfunctions read
\begin{equation}
\Phi_\alpha(\boldsymbol{r},s) = \varphi_{n_z}(z,b_z)\varphi_{n_r}^\Lambda(r_\perp,b_\perp)%
\frac{e^{i\Lambda \phi}}{\sqrt{2\pi}} \chi(s),
\end{equation}
where
\begin{align}
\varphi_{n_z}(z,b_z) &=b^{-1/2}_z\varphi_{n_z}(\xi)%
=b^{-1/2}_z\mathcal{N}_{n_z} H_{n_z}(\xi)e^{-\xi^2/2},\\
\varphi_{n_r}^\Lambda(r_\perp,b_\perp) &= b^{-1}_{\perp}\varphi_{n_r}^\Lambda(\eta)%
=b^{-1}_{\perp}\mathcal{N}_{n_r}^{\Lambda}\sqrt{2}\eta^{|\Lambda|/2}
L_{n_r}^{|\Lambda|}(\eta)e^{-\eta/2}.
\end{align}
$H_{n_z}(\xi)$ and $L_{n_r}^{|\Lambda|}(\eta)$ denote the Hermite and associated Laguerre
polynomials, respectively. The normalization factors are
\begin{equation}
\mathcal{N}_{n_z}=(\sqrt{\pi}2^{n_z}n_z!)^{-1/2} \quad \textnormal{and} \quad
\mathcal{N}^\Lambda_{n_r} = (n_r!/(n_r+|\Lambda|)!)^{1/2}.
\end{equation}
The large and small components of a Dirac spinor are expanded independently in terms of the
oscillator eigenfunctions
\begin{align}
f_i(\boldsymbol{r},s,t) &= \frac{1}{\sqrt{2\pi}} \left( \begin{array}{c}
  f_i^+(z,r_\perp) e^{i\Lambda_- \phi} \\ f_i^-(z,r_\perp) e^{i\Lambda_+ \phi}
  \end{array} \right) = \sum_\alpha^{\alpha_{max}}{f_\alpha^{(i)}\Phi_\alpha(\boldsymbol{r},s) }
  \chi_{t_i}(t),\\
g_i(\boldsymbol{r},s,t) &= \frac{1}{\sqrt{2\pi}} \left( \begin{array}{c}
  g_i^+(z,r_\perp) e^{i\Lambda_- \phi} \\ g_i^-(z,r_\perp) e^{i\Lambda_+ \phi}
  \end{array} \right) =
  \sum_{\tilde{\alpha}}^{\tilde{\alpha}_{max}}{g_{\tilde{\alpha}}^{(i)}\Phi_{\tilde{\alpha}}(\boldsymbol{r},s) }
  \chi_{t_i}(t)  .
\end{align}
To avoid the appearance of spurious states, the quantum numbers $\alpha_{max}$
and $\tilde{\alpha}_{max}$ are chosen in such a way that the corresponding major quantum
numbers $N=2n_r+|\Lambda|+n_z$ are not larger than $N_{max}$ and $N_{max}+1$
for the large and small components, respectively.
\subsubsection{The Dirac Hamiltonian}
The matrix elements of the Dirac Hamiltonian for the case of axially symmetric quadrupole
deformation are given by
\begin{align}
\left( \begin{array}{c} \mathcal{A}_{\alpha \alpha^\prime} \\
 \mathcal{C}_{\alpha \alpha^\prime} \end{array} \right) &=
\delta_{\Lambda \Lambda^\prime}\delta_{m_s m_s^\prime}%
\int_{-\infty}^\infty{d\xi\,\varphi_{n^{}_z}(\xi) \varphi_{n_z^\prime}(\xi)\times} \nonumber \\
&\times\int_0^\infty{ d\eta\,\varphi^\Lambda_{n^{}_r}(\eta)\varphi^\Lambda_{n_r^\prime}(\eta)%
\left[M^*(b_z\xi,b_\perp\sqrt{\eta})\pm V (b_z\xi,b_\perp\sqrt{\eta}) \right]}
\end{align}

\begin{align}
&\mathcal{B}_{\alpha \tilde{\alpha}} =
\delta_{\Lambda \tilde{\Lambda}} \delta_{m_s \tilde{m}_s}\delta_{n_r \tilde{n}_r}
(-1)^{1/2-m_s}\frac{1}{b_z}\left( \delta_{n_z \tilde{n}_z-1} \sqrt{\frac{\tilde{n}_z}{2}}
 - \delta_{n_z \tilde{n}_z+1} \sqrt{\frac{n_z}{2}} \right) \nonumber \\
&+ \delta_{\Lambda \tilde{\Lambda}+1} \delta_{m_s \tilde{m}_s-1}\delta_{n_z \tilde{n}_z}
\frac{\mathcal{N}_{n_r}^\Lambda \mathcal{N}_{\tilde{n}_r}^{\tilde{\Lambda}}}{b_\perp}
\int_0^\infty{d\eta e^{-\eta} \eta^{\Lambda-1/2}L_{n_r}^\Lambda(\eta)
\left(\tilde{L}_{\tilde{n}_r}^{\tilde{\Lambda}}(\eta)
  - \Lambda  L_{\tilde{n}_r}^{\tilde{\Lambda}}(\eta)  \right) } \nonumber \\
&+ \delta_{\Lambda \tilde{\Lambda}-1} \delta_{m_s \tilde{m}_s+1}\delta_{n_z \tilde{n}_z}
\frac{\mathcal{N}_{n_r}^\Lambda \mathcal{N}_{\tilde{n}_r}^{\tilde{\Lambda}}}{b_\perp}
\int_0^\infty{d\eta e^{-\eta} \eta^{\Lambda-1/2}L_{n_r}^\Lambda(\eta)
\left(\tilde{L}_{\tilde{n}_r}^{\tilde{\Lambda}}(\eta)
  +\tilde{\Lambda}  L_{\tilde{n}_r}^{\tilde{\Lambda}}(\eta)  \right) }  .
\end{align}
\subsubsection{Coulomb interaction}
The mean-field potential for protons includes the direct Coulomb field
\begin{equation}
V_C (\boldsymbol{r}) = e^2
\int{d^3r^\prime \frac{\rho_p(\boldsymbol{r}^\prime)}{|\boldsymbol{r}-\boldsymbol{r}^\prime  | } }.
\end{equation}
The logarithmic singularity in the integrand at the point $\boldsymbol{r}=\boldsymbol{r}^\prime$
can be eliminated by using the identity~\cite{Vautherin1973_PRC7-296}
\begin{equation}
\triangle_{\boldsymbol{r}^\prime}|\boldsymbol{r}-\boldsymbol{r}^\prime| = \frac{2}{|\boldsymbol{r}-\boldsymbol{r}^\prime|},
\end{equation}
that, together with an integration by parts, gives
\begin{equation}
V_C(\boldsymbol{r}) = \frac{e^2}{2}
\int{d^3r^\prime |\boldsymbol{r}-\boldsymbol{r}^\prime  |
\triangle_{\boldsymbol{r}^\prime} \rho_p(\boldsymbol{r}^\prime)  }.
\end{equation}
After integrating over the azimuthal angle $\phi$, one obtains the
following expression
\begin{equation}
V_C(r_\perp,z) = 2e^2\int_0^\infty{ r_\perp^\prime dr_\perp^\prime%
\int_{-\infty}^\infty{dz^\prime d(r_\perp,z) E\left(\frac{4r_\perp r_\perp^\prime}{d(r_\perp,z)}\right)%
\triangle \rho_p(r_\perp^\prime,z^\prime)
}
},
\end{equation}
with $d(r_\perp,z)= \sqrt{(z-z^\prime)^2+(r_\perp+r_\perp^\prime)^2}$. The complete elliptic integral
of the second kind is approximated by the standard polynomial formula~\cite{Abramowitz1970}.

\subsubsection{Klein-Gordon equations}
In the case of axial symmetry the Helmholtz equations for the meson fields $\phi$ = $\sigma$, $\omega$, $\rho$ read
\begin{equation}
\left( -\frac{\partial^2}{\partial r_\perp^2}
- \frac{2}{r_\perp}\frac{\partial}{\partial r_\perp} + m_\phi^2\right)
\phi(r_\perp,z) = s_\phi(r_\perp ,z).
\end{equation}
The solution is obtained by expanding the fields in a complete
set of basis states:
\begin{equation}
\label{Eq:KG-expansion-axial}
\phi(z,r_\perp) = \sum^{N_B}_{n_zn_r}\phi_{n_z n_r}\varphi_{n_z}(z,b_z)\varphi^0_{n_r}(r_\perp,b_\perp).
\end{equation}
It is convenient to use the same deformation parameter $\beta_0$ and oscillator
frequency $\hbar \omega_0$ as for the nucleon wave functions.
Inserting the expansion into the Klein-Gordon equation, one obtains an
inhomogeneous set of linear equations
\begin{equation}
\label{Eq:KG-axial}
\sum_{n_z^\prime n_r^\prime}^{N_B} { \mathcal{H}_{n_z n_r n_z^\prime n_r^\prime}
\phi_{n_z^\prime n_r^\prime}  } = s_{n_z n_r}^\phi,
\end{equation}
with the matrix elements
\begin{align}
\mathcal{H}_{n_z n_r n_z^\prime n_r^\prime} &= \delta_{n_r n_r^\prime}
\delta_{n_z n_z^\prime} \left(  \frac{1}{b_z^2} (n_z + \frac{1}{2}) +
 \frac{1}{b_\perp^2} ( 2n_r +1)+ m_\phi^2  \right) \nonumber \\
 &-\frac{1}{2b_z^2} \delta_{n_r n_r^\prime} \left(
 \sqrt{(n_z+1)n_z^\prime}\delta_{n_z n_z^\prime-2}
 + \sqrt{n_z(n_z^\prime+1)}\delta_{n_z n_z^\prime+2} \right) \nonumber \\
 &+\frac{1}{b_\perp^2} \delta_{n_z n_z^\prime} \left( n_r^\prime \delta_{n_r n_r^\prime-1}
 +n_r \delta_{n_r n_r^\prime+1} \right).
\end{align}
The set of equations (\ref{Eq:KG-axial}) is solved by inversion.
\subsubsection{The pairing matrix elements}
\label{subsubsec:pairing-axial}
Although the total angular momentum $J$ is no longer a good
quantum number, $\Omega_{tot}=0$ is still valid in the pairing channel.
For the matrix element of the pairing interaction Eq.~(\ref{Eq:Vpp})
the operator $(1-P^\sigma)/2$ projects onto the $S=0$ spin-singlet product space
\begin{align}
|\gamma \bar{\delta}\rangle_{S=0} = \frac{1}{\sqrt{2}}
\varphi_{n_z^\gamma}(z_1,b_z)\varphi_{n_z^\delta}(z_2,b_z)
&\varphi_{n_r^\gamma}^{\Lambda^\gamma}(r_{\perp,1},b_\perp)
\varphi_{n_r^\delta}^{\Lambda^\delta}(r_{\perp,2},b_\perp) \times \nonumber \\
&\times \frac{1}{2\pi} e^{i\Lambda^\gamma (\phi_1- \phi_2)}
\delta_{\Lambda^\gamma \Lambda^\delta}\delta_{m_s^\gamma m_s^\delta}.
\end{align}
Since the separable pairing interaction is expressed in terms of
the center-of-mass coordinate $\boldsymbol{R}$ and the relative coordinate
$\boldsymbol{r}$ of a pair, the two-particle wave function is transformed to the center-of-mass frame:
\begin{equation}
|n_z^\gamma n_z^\delta \rangle = \sum_{N_z n_z}{M^{N_z n_z}_{n_z^\gamma\,n_z^\delta}
|N_z n_z \rangle },
\end{equation}
\begin{equation}
|n_r^\gamma \Lambda^\gamma n_r^\delta \Lambda^\delta \rangle = \sum_{N_r \Lambda}{
\sum_{n_r \lambda}{ M^{N_r\Lambda~n_r \lambda}_{n_r^\gamma \Lambda^\gamma\,n_r^\delta \Lambda^\delta}%
| N_r \Lambda n_r \lambda \rangle } },
\end{equation}
where $\displaystyle M^{N_z n_z}_{n_z^\gamma\,n_z^\delta}$ and
$\displaystyle  M^{N_r\Lambda~n_r \lambda}_{n_r^\gamma \Lambda^\gamma\,n_r^\delta \Lambda^\delta}$
denote the one- and two-dimensional Talmi-Moshinski transformation brackets, respectively \cite{Talman1970_NPA141-273,Chaos-Cador2004_IJQC97-844}.
The selection rule $\Lambda^\gamma+\Lambda^\delta=\Lambda+\lambda$ yields the
constraint $\Lambda+\lambda=0$. The pairing matrix element can be expressed as a
sum over the complete set of quantum numbers $N_z$, $n_z$, $N_r$, $n_r$, $\Lambda$, $\lambda$,
$N_z^\prime$, $n_z^\prime$, $N_r^\prime$, $n_r^\prime$, $\Lambda^\prime$ and
$\lambda^\prime$. However,
the factor $\delta(\boldsymbol{R}-\boldsymbol{R}^\prime)$ in the pairing interaction restricts this
sum to $N_z=N_z^\prime$, $N_r=N_r^\prime$, $\lambda=\lambda^\prime$,
$\Lambda=\Lambda^\prime$. Furthermore, the angular part of the integral over the relative coordinate vanishes for $\lambda \neq 0$, that is, $\Lambda=\lambda=0$. Thus
the matrix element of the separable force Eq.~(\ref{Eq:Vpp}) in the axially deformed oscillator basis can be written as a sum over two quantum numbers: $N_z$ and $N_r$,
\begin{equation}
\langle \alpha \bar{\beta} | V^{pp} |\gamma \bar{\delta} \rangle_a =
 -G \sum_{N_z N_r}{ W_{\alpha \bar{\beta}}^{N_z N_r*}
 W_{\gamma \bar{\delta}}^{N_z N_r}},
\end{equation}
The summation over the
quantum numbers $n_z$, $n_r$, $n_z^\prime$ and $n_r^\prime$ vanishes because of
the conservation of the harmonic oscillator quanta
 \begin{align}
n_z &= n_{z_1} + n_{z_2} - N_z,\\
n_r &= n_{r_1}+n_{r_2} + |\Lambda_1| - N_r.
\end{align}
The single-particle matrix element $W_{\gamma \bar{\delta}}^{N_z N_r}$ can be factorized
into two one-dimensional integrals
\begin{equation}
W_{\gamma \bar{\delta}}^{N_z N_r} =W^{N_z}_{\gamma \bar{\delta}} W^{N_r}_{\gamma \bar{\delta}}.
\end{equation}
For $W_{\gamma \bar{\delta}}^{N_z}$ we find
\begin{equation}
W_{\gamma \bar{\delta}}^{N_z} = \frac{1}{b^{1/2}_z}M^{N_z n_z}_{n_z^\gamma n_z^\delta } I_{n_z}(\alpha_z)\;,
\end{equation}
where $\alpha_z=a/b_z$. The integral over the relative coordinate
\begin{equation}
I_{n}(\alpha)=\frac{1}{\alpha}\int_{-\infty}^\infty{ \varphi_{n}(\xi) e^{-\xi^2/2\alpha^2}d\xi },
\end{equation}
vanishes for odd values of the quantum number $n$. The integration can be performed
using the generating function for the Hermite polynomials~\cite{Abramowitz1970}
\begin{equation}
 e^{2\xi z-z^2} = \sum_{n=0}^\infty{\frac{1}{n!} H_n(\xi)z^n}.
 \end{equation}
and one obtains
\begin{equation}
\label{Eq:int-z-axial}
I_{n}(\alpha) = \delta_{n,even}\frac{(-1)^{n/2}}{(2\pi)^{1/4}} \frac{\sqrt{n!}}{2^{{n}/2}(n/2)!}
\left(\frac{1}{1+\alpha^2}\right)^{1/2}
\left(\frac{1-\alpha^2}{1+\alpha^2}\right)^{n/2}.
 \end{equation}
The integral over the perpendicular coordinate with $\alpha_\perp=a/b_\perp$
\begin{equation}
W^{N_r}_{\gamma \bar{\delta}} = \frac{1}{b_\perp}%
M^{N_r0~n_r0}_{n_r^\gamma \Lambda^{\gamma} n_r^\delta -\Lambda^\gamma}%
\frac{1}{\alpha_\perp^2} \int_0^\infty{\varphi_{n_r}^0(\eta) e^{-\eta^2/4\alpha_\perp^2}\eta d\eta },
\end{equation}
can be evaluated using the generating function for the Laguerre polynomials~\cite{Abramowitz1970}
 \begin{equation}
 \frac{e^{-\eta z/(1-z)}}{1-z} = \sum_{n=0}^\infty{ \frac{1}{n!} L_n(\eta) z^n },
 \end{equation}
 with the resulting expression
  \begin{equation}
 W^{N_r}_{\gamma \bar{\delta}} =\frac{1}{b_\perp}
 M^{N_r0~\,n\,0}_{n_r^\gamma \Lambda^\gamma n_r^\delta-\Lambda^\delta}
 \frac{1}{(2\pi)^{1/2}} \frac{1}{1+\alpha^2_\perp}
 \left(\frac{1-\alpha^2_\perp}{1+\alpha^2_\perp} \right)^{n_r}.
 \end{equation}

\subsection{Nuclei with triaxial quadrupole shapes}
\label{subsection:triaxial}
The Dirac single-nucleon spinors are expanded in a basis of eigenfunctions of a three-dimensional
harmonic oscillator in Cartesian coordinates~\cite{Koepf1988_PLB212-397,PENG2008_PRC78-024313}:
\begin{equation}
V(x,y,z) = \frac{1}{2}m\omega_x^2 x^2 +\frac{1}{2}m\omega_y^2 y^2
              + \frac{1}{2}m\omega_z^2 z^2.
\end{equation}
Because of volume conservation, the three oscillator frequencies can be expressed in terms
of the basis deformation parameters $\beta_0$ and $\gamma_0$,
\begin{align}
\hbar \omega_x &= \hbar\omega_0 e^{-\sqrt{\frac{5}{4\pi}}\beta_0 \cos{(\gamma_0-2\pi/3)}},\\
\hbar \omega_y &= \hbar\omega_0 e^{-\sqrt{\frac{5}{4\pi}}\beta_0 \cos{(\gamma_0+2\pi/3)}},\\
\hbar \omega_z &= \hbar\omega_0 e^{-\sqrt{\frac{5}{4\pi}}\beta_0 \cos{\gamma_0}}.
\end{align}
The corresponding oscillator lengths are $b_\mu = \sqrt{\hbar/m\omega_\mu}$ and, because of volume
conservation, $b_xb_xb_z=b_0^3$. The basis is completely determined by the three constants
$\hbar \omega_0$, $\beta_0$, and  $\gamma_0$. The basis states are products of
three one-dimensional harmonic oscillator eigenfunctions and the spin factor:
\begin{equation}
\Phi_\alpha(\boldsymbol{r}; m_s) = \varphi_{n_x}(x,b_x)  \varphi_{n_y}(y,b_y)  \varphi_{n_z}(z,b_z)\chi_{m_s},
\end{equation}
The one-dimensional oscillator eigenfunctions read
\begin{equation}
\varphi_{n_\mu}(x_\mu,b_\mu) = b_\mu^{-1/2}\varphi_{n_\mu}(\xi_\mu)= b_\mu^{-1/2}\mathcal{N}_{n_\mu} H_{n_\mu}(\xi_\mu)
e^{-\xi_\mu^2/2} \quad (\mu \equiv x,y,z),
\end{equation}
where $\xi_\mu$ denotes the ratio between the Cartesian coordinate and the corresponding
oscillator length $b_\mu = \sqrt{\hbar/m\omega_\mu}$ . The normalization factor reads
\begin{equation}
\mathcal{N}_n = (\sqrt{\pi}2^n n!)^{-1/2},
\end{equation}
and the Hermite polynomial $H_n(\xi)$ is defined in Ref.~\cite{Abramowitz1970}.

The spatial part of the wave function is labeled by the quantum numbers $\alpha=\{n_x,n_y,n_z\}$.
For each combination of these quantum numbers the spin part
is chosen in such a way that the basis state is
an eigenfunction of the $x$-simplex operator $\hat{S}_x = \hat{P} e^{-i\pi J_x}$, where
$\hat{P}$ denotes the parity operator. The positive and negative $x$-simplex operator~\cite{PENG2008_PRC78-024313} eigenstates read
\begin{align}
| n_x n_y n_z; i=+\rangle &= | n_x n_y n_z \rangle \frac{i^{n_y}}{\sqrt{2}} \left[
|\uparrow \rangle - (-1)^{n_x} | \downarrow \rangle \right] , \\
| n_x n_y n_z; i=-\rangle &= | n_x n_y n_z \rangle (-1)^{n_x+n_y+1}\frac{i^{n_y}}{\sqrt{2}}%
\left[|\uparrow \rangle + (-1)^{n_x} | \downarrow \rangle \right] ,
\end{align}
and are related by the time-reversal operator
\begin{equation}
| n_x n_y n_z; i=- \rangle = \hat{T} | n_x n_y n_z; i=+ \rangle .
\end{equation}
For a Dirac spinor with a positive simplex eigenvalue, the large component corresponds
to positive eigenvalues and the small component to negative eigenvalues. The large and
small components are expanded in terms of the basis states
\begin{equation}
f_i(\boldsymbol{r},\pm) = \sum_{\alpha}^{\alpha_{max}}{f_i^\alpha \Phi_\alpha(\boldsymbol{r};\pm)  },\quad
g_i(\boldsymbol{r},\pm) = \sum_{\tilde{\alpha}}^{\tilde{\alpha}_{max}}{g_i^{\tilde{\alpha}}
\Phi_{\tilde{\alpha}}(\boldsymbol{r};\pm)  }.
\end{equation}
In the present implementation of the model parity is also conserved, thus allowing
a further reduction of the basis to four simplex-parity blocks.
To avoid the appearance of spurious states, the quantum numbers $\alpha_{max}$
and $\alpha_{max}$ are chosen in such a way that the corresponding major quantum number
$N=n_x+n_y+n_z$ does not exceed $N_{max}$ and $\tilde{N}_{max}=N_{max}+1$ for large and
small components, respectively.
\subsubsection{The Dirac Hamiltonian}
In addition to x-simplex and parity, one assumes that the densities and fields
are reflection-symmetric with respect to the $yz$, $xz$ and $xy$ planes
\begin{align}
\label{Eq:D2}
V(-x,y,z)=V(x,y,z), \\ V(x,-y,z)=V(x,y,z), \\ V(x,y,-z)=V(x,y,z).
\end{align}
The matrix elements of the Dirac Hamiltonian in Cartesian coordinates are given by
\begin{align}
\left( \begin{array}{c}
\mathcal{A}_{\alpha \alpha^\prime} \\ \mathcal{C}_{\alpha \alpha^\prime}
\end{array} \right)
&= (-1)^{(n_y+n_y^\prime)/2}
\delta_{n_x+n_x^\prime,even}\delta_{n_y+n_y^\prime,even}\delta_{n_z+n_z^\prime,even}
\times \nonumber \\ &\times
\langle n_x n_y n_z | M^*(x,y,z) \pm V(x,y,z) | n_x^\prime n_y^\prime n_z^\prime \rangle
\end{align}
\begin{align}
\mathcal{B}_{\alpha \tilde{\alpha}} &=
\frac{1}{\sqrt{2}b_x} \delta_{n_y \tilde{n}_y}\delta_{n_z \tilde{n}_z} (-1)^{n_x+n_y}
\left[  \sqrt{n_x} \delta_{n_x \tilde{n}_x+1} - \sqrt{\tilde{n}_x}\delta_{n_x \tilde{n}_x-1}
\right] \nonumber \\
&+ \frac{1}{\sqrt{2}b_y} \delta_{n_x \tilde{n}_x}\delta_{n_z \tilde{n}_z} (-1)^{n_y}
\left[  \sqrt{n_y} \delta_{n_y \tilde{n}_y+1} + \sqrt{\tilde{n}_y}\delta_{n_y \tilde{n}_y-1}
\right] \nonumber \\
&+ \frac{1}{\sqrt{2}b_z} \delta_{n_x \tilde{n}_x}\delta_{n_y \tilde{n}_y} (-1)^{n_x+n_y}
\left[  \sqrt{n_z} \delta_{n_z \tilde{n}_z+1} - \sqrt{\tilde{n}_z}\delta_{n_z \tilde{n}_z-1}
\right].
\end{align}
\subsubsection{Coulomb interaction}
The direct Coulomb potential is given by the three-dimensional integral
\begin{equation}
U_c(\boldsymbol{r}) = e^2 \int{d^3r^\prime
\frac{\rho_p(\boldsymbol{r}^\prime)}{|\boldsymbol{r}-\boldsymbol{r}^\prime|} }.
\end{equation}
Although a direct integration of the Poisson equation can easily be performed
in calculations when the problem is characterized by spherical or axial symmetry,
the number of mesh points on the three dimensional grid can easily become prohibitively large.
In this case the Poisson equation can be solved using the conjugate gradient method~\cite{Bonche2005_CPC171-49}, or the Coulomb
interaction is replaced by an integral over Gaussian interactions~\cite{Girod1983_PRC27-2317,Stoitsov2005_CPC167-43}.
The DIRHB triaxial code employs a method based on the Coulomb Green's function,
that is also used in the non-relativistic HFODD code~\cite{Dobaczewski1997_CPC102-166}.
The direct Coulomb potential can be expressed through the Dirichlet Green's function
$G_D(\boldsymbol{r},\boldsymbol{r}^\prime)$ in the following way~\cite{Jackson1962}
\begin{equation}
U_c(\boldsymbol{r})  = e^2 \int_V{d^3r^\prime G_D(\boldsymbol{r},\boldsymbol{r}^\prime)
\rho_p(\boldsymbol{r}^\prime)}
-\frac{e^2}{4\pi} \oint_S{d^2s^\prime \frac{\partial G_D(\boldsymbol{r},\boldsymbol{r}^\prime)}{\partial n^\prime}
U_c(\boldsymbol{r}^\prime) }.
\end{equation}
The first term is the volume integral over an arbitrary closed volume, whereas the
second integral is evaluated over the surface enclosing this volume. The Dirichlet
Green's function presents a solution of the Poisson equation for a point charge,
and vanishes at the surface.
The normal derivative in the surface term is calculated with respect to the outward
direction perpendicular to the surface. The method proposed in
Ref.~\cite{Dobaczewski1997_CPC102-166}
adopts the volume in the form of the parallelepiped
\begin{equation}
-D_x \le x \le D_x, \quad -D_y \le y \le D_y, \quad -D_z \le z \le D_z.
\end{equation}
The Dirichlet Green's function can be expressed in a separable form as
\begin{equation}
G_D(\boldsymbol{r},\boldsymbol{r}^\prime) = \frac{4\pi}{D_xD_yD_z}
\sum_{j_xj_yj_z}{
\frac{f(J_x x)f(J_y y)f(J_z z)f(J_x x^\prime)f(J_y y^\prime)f(J_z z^\prime)}{J_x^2+J_y^2+J_z^2} },
\end{equation}
where the functions $f$ ensure the Dirichlet boundary conditions,
\begin{equation}
f(J_i x_i) = \left\{ \begin{array}{ll}
\cos{(J_i x_i)} & \textnormal{for}\; j_i \; \textnormal{even}, \\
\sin{(J_i x_i)} & \textnormal{for}\; j_i \; \textnormal{odd},
\end{array} \right. \quad \textnormal{with} \quad
J_i = \frac{(j_i+1)\pi}{2D_i}.
\end{equation}
The Coulomb potential on the surface of the parallelepiped is approximated by the
multipole expansion
\begin{equation}
U_c(\boldsymbol{r}) = \sum_{\lambda \mu}{\frac{4\pi e^2}{(2\lambda+1)r^{2\lambda+1}}
Q^{(p)}_{\lambda \mu}r^\lambda Y_{\lambda \mu}(\theta, \phi)  }.
\end{equation}
In practical calculations it suffice to retain the monopole, quadrupole and
hexadecupole term in the multipole expansion.
\subsubsection{The Helmholtz equation}
In Cartesien coordinates the Helmholtz equations for the meson fields $\phi$ = $\sigma$, $\omega$, $\rho$ read
\begin{equation}
\left( -\frac{\partial^2}{\partial x^2}  -\frac{\partial^2}{\partial y^2}
-\frac{\partial^2}{\partial
z^2} + m_\phi^2\right)
\phi(x,y,z) = s_\phi(x,y,z).
\end{equation}
The solution is obtained by expanding the fields in a complete
set of basis states
\begin{equation}
\label{Eq:KG-expansion-cartesian}
\phi(x,y,z) = \varphi_{n_x}(x,b_x)\varphi_{n_y}(y,b_y) \varphi_{n_z}(z,b_z).
\end{equation}
Also in this case the same deformation parameter $\beta_i$  is used as for the nucleon wave functions.
Inserting the expansion into the Klein-Gordon equation, one obtains an
inhomogeneous set of linear equations
\begin{equation}
\label{Eq:KG-cartesian}
\sum_{n_x^\prime n_y^\prime n_z^\prime}^{N_B} {
\mathcal{H}_{n_x n_y n_z n_x^\prime n_y^\prime n_z^\prime}
\phi_{n_x^\prime n_y^\prime n_z^\prime}  } = s_{n_x n_y n_z}^\phi,
\end{equation}
with the matrix elements given by
\begin{align}
\mathcal{H}_{n_x n_y n_z n_x^\prime n_y^\prime n_z^\prime} &=
-\frac{1}{b_x^2} \left[ \sqrt{(n_x+1)n_x^\prime}\delta_{n_x n_x^\prime-2}
   +  \sqrt{n_x (n_x^\prime+1)}\delta_{n_x n_x^\prime+2} \right] \delta_{n_y n_y^\prime}
   \delta_{n_z n_z^\prime} \nonumber \\
&-\frac{1}{b_y^2} \left[ \sqrt{(n_y+1)n_y^\prime}\delta_{n_y n_y^\prime-2}
   +  \sqrt{n_y (n_y^\prime+1)}\delta_{n_y n_y^\prime+2} \right] \delta_{n_x n_x^\prime}
   \delta_{n_z n_z^\prime} \nonumber \\
&-\frac{1}{b_z^2} \left[ \sqrt{(n_z+1)n_z^\prime}\delta_{n_z n_z^\prime-2}
   +  \sqrt{n_z (n_z^\prime+1)}\delta_{n_z n_z^\prime+2} \right] \delta_{n_x n_x^\prime}
   \delta_{n_y n_y^\prime} \nonumber \\
&+ \left[ \frac{n_x+\frac{1}{2}}{b_x^2} +  \frac{n_y+\frac{1}{2}}{b_y^2}
+ \frac{n_z+\frac{1}{2}}{b_z^2}  + m_\phi^2\right] \delta_{n_x n_x^\prime}
\delta_{n_y n_y^\prime} \delta_{n_z n_z^\prime} .
\end{align}
The set of equations (\ref{Eq:KG-cartesian}) is solved by inversion.
\subsubsection{Pairing matrix elements}
The antisymmetric matrix elements of the pairing interaction Eq.~(\ref{Eq:Vpp})
can be separated into a product of spin and  coordinate space factors.
The operator $(1-P^\sigma)/2$ projects onto the $S=0$ spin-singlet product state
\begin{align}
|\gamma \bar{\delta} \rangle_{S=0} &= -| \bar{\delta}\gamma \rangle_{S=0}
=\frac{1}{2}i^{n_y^\gamma+n_y^\delta}(-1)^{n_y^\delta+1} \nonumber \\
&\times \delta_{n_x^\gamma+n_x^\delta,\textnormal{even}} \left[ |\uparrow \downarrow\rangle
-|\downarrow \uparrow \rangle \right] |n^\gamma n^\delta \rangle,
\end{align}
and the problem reduces to the calculation of the spatial part of the matrix element.
This integral is separable in the $x$-, $y$- and $z$-coordinates. As an example, the $x$ component reads
\begin{align}
V_x &= \int{\varphi_{n_x^\alpha}(x_1,b_x) \varphi_{n_x^\beta}(x_2,b_x) P(x)\delta(X-X^\prime)
P(x^\prime) } \nonumber \\
&\times \varphi_{n_x^\gamma}(x_1^\prime,b_x) \varphi_{n_x^\delta}(x_2^\prime,b_x)
dx_1 dx_2 dx_1^\prime dx_2^\prime.
\end{align}
This integral is evaluated by making use of the 1D Talmi-Moshinsky
transformation and the generating function for the Hermite polynomials (see Sect.~\ref{subsubsec:pairing-axial}). One finds that the pairing matrix element,
\begin{equation}
\langle \alpha \bar{\beta} | V^{pp} | \gamma \bar{\delta} \rangle_a =
-G \sum_{N_x=0}^{N_x^0}{\sum_{N_y=0}^{N_y^0}{\sum_{N_z=0}^{N_z^0}{
\left(W_{\alpha \bar{\beta}}^{N_xN_yN_z}  \right)^*
W_{\gamma \bar{\delta}}^{N_xN_yN_z}  }    }     } ,
\end{equation}
can be represented as a sum of separable terms in a 3D Cartesian harmonic oscillator
basis with the single-particle matrix elements
\begin{equation}
W_{\gamma \bar{\delta}}^{N_xN_yN_z}  = \delta_{n_x^\gamma + n_x^\delta, even}
i^{n_y^\gamma-n_y^\delta} W_{n_x^\gamma n_x^\delta}^{N_x}
W_{n_y^\gamma n_y^\delta}^{N_y} W^{N_z}_{n_z^\gamma n_z^\delta}.
\end{equation}
The factors $W_{n_1 n_2}^{N_\mu}$ are given by
\begin{equation}
W^{N_\mu}_{n_\mu^\gamma n_\mu^\delta} = \frac{1}{b^{1/2}_\mu}M^{n\,N_\mu}_{n_\mu^\gamma n_\mu^\delta} I_{n}(\alpha_\mu) \quad \textnormal{with} \quad
n=N_\mu-n_\mu^\gamma-n_\mu^\delta .
\end{equation}
where $M^{n\,N}_{n_\gamma n_\delta}$ denotes the 1D Talmi-Moshinsky bracket, $\alpha_\mu=a/b_\mu$, and the integral $I_n(\alpha)$ is given
in Eq.~(\ref{Eq:int-z-axial}).

Because of reflection symmetry (\ref{Eq:D2}) of the potentials and corresponding densities, only
those matrix elements with even values for $n_x^\gamma+n_x^\delta$ , $n_y^\gamma+n_y^\delta$,
and $n_z^\gamma+n_z^\delta$ contribute to
$P^{N_xN_yN_z}=\frac{1}{2}\textnormal{Tr}(W^{N_xN_yN_z}\kappa)$ and, therefore, $N_x$, $N_y$, and $N_z$ have to be even. As a consequence, only matrix elements of $\Delta_{\gamma \bar{\delta}}$ with even values for $n_x^\gamma+n_x^\delta$, $n_y^\gamma+n_y^\delta$, and $n_z^\gamma+n_z^\delta$ do not vanish.
\section{Structure of the {\tt DIRHB} program package}
\label{sec:program-structure}
All three codes included in the {\tt DIRHB} package
({\tt DIRHBS, DIRHBZ and DIRHBT}) consist of a Fortran program and two additional files: {\tt DIRHB.PAR} and {\tt DIRHB.DAT}. The file
{\tt DIRHB.PAR} contains the relevant information about the dimension of the arrays,
depending on the number of oscillator shells selected for the expansion of nucleon spinors
({\tt N0FX}) and boson fields ({\tt N0BX}), and the number of Gaussian mesh points.
The file {\tt DIRHB.DAT} includes the data for the specific nucleus being calculated.
The main program calls various subroutines that read the data and perform the computation.
The execution essentially consists of three parts. The first part uses
the file {\tt DIRHB.DAT} to start the program, initializes and generates
all the relevant information. It uses the subroutines
{\tt DEFAULT}, {\tt READER}, {\tt FORCES}, {\tt GAUSS}, {\tt BASE},
{\tt PREP},  {\tt INOUT}, {\tt DINOUT}, {\tt START}, {\tt GAUPOL}, {\tt SINGF} and
{\tt SINGD}, {\tt GREEMES}, and {\tt GREECOU}.

The second part of the code carries out the self-consistent computation.
The iterative procedure is performed by the
subroutine {\tt ITER}.
\begin{table}
\centering
\caption{Program structure of the {\tt DIRHB} computer codes.}
\begin{tabular}{ll}
\hline
Subroutine & Task  \\ \hline
-main  & \\
\hspace{1cm}-default & Initializes all variables. \\
\hspace{1cm}-reader & Reads parameters from the input data file {\tt dirhb.dat} . \\
\hspace{1cm}-prep & Prepares variables according to the input data file. \\
\hspace{1cm}-inout and dinout & Reads initial fields ({\tt inin}=0) and pairing 
                    tensor ({\tt inink}=0).  \\
\hspace{1cm}-start & Sets up the initial Woods-Saxon fields ({\tt inin}=1) and \\ 
                   & pairing tensor ({\tt inink}=1).\\
\hspace{1cm}-base & Constructs  the configuration space and the arrays of
                                 quantum  \\  & numbers. \\
\hspace{1cm}-gaupol & Calculates the HO basis functions at the Gaussian mesh points. \\
\hspace{1cm}-singf & Calculates  the single-particle matrix elements. \\
\hspace{1cm}-singd & Calculates the pairing matrix elements .\\
\hspace{1cm}-greecou & Calculates  the Coulomb Green's function. \\
\hspace{1cm}-greemes & Calculates  the meson Green's function.\\
\hspace{1cm}-iter & Main iteration loop, repeated until convergence is reached. \\
\hspace{2cm}-gamma& Calculates  the Dirac Hamiltonian matrix elements. \\
\hspace{2cm}-delta&  Calculates and stores matrix elements of  the pairing field. \\
\hspace{2cm}-dirhb& Diagonalizes the RHB equation \\
\hspace{2cm}-denssh& Calculates the density matrix and pairing tensor in the \\
            & oscillator space. \\
\hspace{2cm}-densit&   Transforms the densities into coordinate space. \\
\hspace{2cm}-gdd&  Calculates the density dependent couplings in coordinate\\
  & space. \\
\hspace{2cm}-field&  Calculates the fields in coordinate space. \\
\hspace{2cm}-coulom& Calculates the Coulomb potential in coordinate space. \\
\hspace{2cm}-expect&  Calculates various expectation values. \\
\hspace{2cm}-cstrpot&  Calculates the constraining potential \\
           & (relevant only for the axial and the triaxial code). \\
\hspace{2cm}-poten&  Calculates the potentials in coordinate space. \\
\hspace{2cm}-itestc&  Verifies whether the  convergency criterion is satisfied. \\
\hspace{1cm}-canon&  Constructs  the canonical basis. \\
\hspace{1cm}-centmas& Calculates the center-of-mass correction. \\
\hspace{1cm}-resu & Prints out the  quasiparticle properties. \\
\hspace{1cm}-inout & Stores the final potentials for future use.  \\
\hspace{1cm}-dinout&  Stores the pairing matrix elements for future use.\\
\hspace{1cm}-plot&  Prints out the densities in coordinate space.\\
\hline
\end{tabular}
\end{table}
When the parameter {\tt IBROYD} is set to 0, the intermediate solutions are combined
using the linear mixing procedure.
In this case the program runs interactively.
After three iterations the user has to input the number of iterations, and the value of
the parameter {\tt xmix} that determines the amount of mixing between the new
matrix elements and those calculated in the previous iteration
\begin{equation}
\phi_{n+1}(\boldsymbol{r}) = x_{mix}\phi_{n+1}(\boldsymbol{r})+(1-x_{mix})\phi_n(\boldsymbol{r}).
\end{equation}
For example {\tt maxit} can be set to 100, and the value of {\tt xmix} to 0.2. For the parameter
{\tt xmix} to change automatically during the iteration procedure, the value has to be input
with a negative sign, that is,  -100
instead of 100. Setting the value of {\tt maxit} to 0 stops the iteration immediately.
When the parameter {\tt IBROYD} (predefined in the subroutine {\tt DEFAULT})
is set to 1, the intermediate solutions are combined using the Broyden mixing procedure~\cite{Baran2008_PRC78-014318}.
In this mode the program runs automatically with the mixing parameter {\tt xmix = 0.5}.
At each iteration step the program prints out the
iteration number, the current level of accuracy ({\tt si}) that corresponds to the maximal difference
between the matrix elements calculated in the previous and current iteration,
the current mixing parameter
between the previous and fields from the present iteration ({\tt xmix}), the total energy per
particle\footnote{to speed up the calculation, the center-of-mass
correction is computed only after convergence is achieved}
and the current value of the mass distribution $rms$ radius. For the {\tt DIRHBZ}
code the printout also contains the deformation parameter $\beta$ and, correspondingly, both deformation
parameters $\beta$ and $\gamma$ for the {\tt DIRHBT} code. The iteration procedure
stops when, for all mean-field and pairing matrix elements, the changes
between two consecutive iteration steps become smaller than the parameter {\tt epsi}, predefined in
the  subroutine {\tt DEFAULT}.

The third part of the code performs the remaining computations after convergence
is reached. The quasiparticle states and the single-particle states
in the canonical basis are printed out, the center-of-mass correction is
evaluated, and various expectation values are computed and printed.
\subsection{Input and output data}
\label{subsec:input}
The input data include:
\begin{itemize}
\item Number of oscillator shells for fermions and bosons ({\tt n0f}, {\tt n0b}). These
   numbers must not exceed the values of the parameters {\tt n0fx} and {\tt n0bx} in the file
   {\tt dirhb.par}.
\item $\beta$-deformation parameter of the harmonic oscillator basis ({\tt beta0} for the axial and triaxial codes).
\item $\gamma$-deformation parameter of the harmonic oscillator basis ({\tt gamma0} for the triaxial code).
\item $\beta$-deformation parameter for the initial Woods-Saxon potential
({\tt betai}, only for the axial and triaxial codes).
\item $\gamma$-deformation parameter of the initial Woods-Saxon potential
({\tt gammai}, only for the triaxial code).
\item The starting parameter for the potential ({\tt inin}).
If the parameter {\tt inin} is set to 1, the code starts
from a default Woods-Saxon potential predefined in the code. If the parameter {\tt inin} is set to 0,
the initial potential is read from the file {\tt dirhb.wel}.
\item The starting parameter for the potential and
pairing field ({\tt inink}). If {\tt inink} is set to 1, the code starts
with the diagonal pairing field with equal matrix elements {\tt delta0}. If {\tt in ink} is set to 0,
the initial pairing matrix elements are read from the file {\tt dirhb.del}.
\item Neutron and proton initial pairing gaps ({\tt delta0}).
\item The nuclide to be computed: the element name ({\tt nucnam}) followed by
the mass number ({\tt nama}).
\item Acronym of the parameter set of the selected energy density functional ({\tt parname}).
\end{itemize}
The following parameters, used to control the constrained calculation, are used only
by the {\tt DIRHBZ} and {\tt DIRHBT} codes. In the case of axial symmetry only $\beta$ is constrained.
\begin{itemize}
\item The quadrupole constraint control parameter {\tt icstr}. If {\tt icstr} is set to 0, the
quadrupole constraint is not included, and the parameters {\tt betac}, {\tt gammac} and
{\tt cqad} are not used. If {\tt icstr} is set to 1, then {\tt betac} and {\tt gammac} denote the
constrained value of the quadrupole deformation in the $\beta - \gamma$ plane.
\item Constrained value of the $\beta$-deformation parameter ({\tt betac}).
\item Constrained value of the $\gamma$-deformation parameter ({\tt gammac}).
\item Stiffness constant for the quadrupole constraint ({\tt cqad}).
\end{itemize}

The output of the calculation is rather simple and we only briefly describe
the different sections of the output file {\tt dirhb.out}.
\begin{itemize}
\item {\it Header}: The title of the code, date and time of execution, and the
name, mass, neutron and proton number of the nucleus, the acronym of the effective interaction used in the
calculation.
\item {\it Input data}: Summary of the input data.
\item {\it Force}: The acronym and parameters of the effective interaction used in the
particle-hole and particle-particle channels.
\item {\it Numerical}: Summary of the numerical parameters and options.
\item {\it Iterations}: Brief information for each iteration step.
\item {\it Canonical basis}: Quantum numbers of the harmonic oscillator basis state that corresponds to
the largest component of the canonical wave function, contribution of this basis state in the
canonical wave function,  single-particle energies, pairing gaps and occupation
probabilities in the canonical basis.
\item {\it Quasiparticle basis}: Quantum numbers of the harmonic oscillator basis state that corresponds to
the largest component of the quasiparticle wave function,
quasiparticle energies, norms of the $U$ and $V$ coefficients. If the norm of the coefficient $V$
is larger than $0.5$ the state is labeled as a hole state ({\it h}), otherwise it is
labeled as a particle state ({\it p}).
\item {\it Observables}: Expectation values of various observables calculated for the
final RHB state.
\end{itemize}
The trace of the vector/scalar density is calculated by integrating the corresponding density.
The charge radius is calculated using the simple formula:
\begin{equation}
r_c = \sqrt{r_p^2+0.64} \quad \textnormal{(fm)} ,
\end{equation}
where $r_p$ denotes the $rms$ radius of the proton density distribution.
The term $0.64$ fm$^2$ accounts for the finite size of the proton.
The quadrupole moments $Q_{20}^{(n,p)}$ and $Q_{22}^{(n,p)}$ are calculated using
the expressions
\begin{equation}
Q_{20}^{(n,p)} = \sqrt{\frac{5}{16\pi}}\langle 3z^2 -r^2\rangle_{(n,p)},\quad
Q_{22}^{(n,p)} =  \sqrt{\frac{5}{32\pi}}\langle x^2 -y^2\rangle_{(n,p)} .
\end{equation}
The following relations for the quadrupole moments expressed in terms of the
deformation parameters $a_{20}$ and $a_{22}$
\begin{equation}
Q_{20}^{(n,p)}=\frac{3A}{4\pi}R_0^2 a_{20}^{(n,p)},\quad
Q_{22}^{(n,p)} = \frac{3A}{4\pi}R_0^2 a_{22}^{(n,p)} ,
\end{equation}
determine the deformation parameters $\beta$ and $\gamma$:
\begin{equation}
\beta = \sqrt{a_{20}^2 + 2a_{22}^2}, \quad
\gamma = \arctan{\left( \sqrt{2}\frac{a_{22}}{a_{20}} \right)},
\end{equation}
with $R_0=1.2 A^{1/3}$ (fm). The sign convention is that of  Ref.~\cite{Ring1980}.
For the case of axial symmetry,  the deformation parameter $\beta$ is determined
by the quadrupole moment $Q_{20}$:
\begin{equation}
Q_{20} = \sqrt{\frac{9}{5\pi}}AR_0^2 \beta.
\end{equation}
Positive $\beta$ values correspond to the $\gamma=0^0$ axis (prolate shapes), and
negative values to the $\gamma=180^0$ axis (oblate shapes).
The hexadecapole moment is computed using the expression
\begin{equation}
H_{n,p}=\langle 8z^4 - 24z^2(x^2+y^2) + 3(x^2+y^2)^2\rangle_{n,p}
\end{equation}

After the self-consistent solution of the RHB equations is reached, the
center-of-mass correction
\begin{equation}
E_{cm} = -\frac{\boldsymbol{P}_{cm}^2}{2AM}
\end{equation}
is subtracted from the total binding energy.
$\boldsymbol{P}_{cm}$ is the total momentum of a nucleus with $A$ nucleons.

The densities in coordinate space are stored in the file {\tt dirhb.plo}. The details
depend on the particular code:
\begin{itemize}
\item {\tt DIRHBS}: prints out the total vector density as a function of the radial coordinate.
\item {\tt DIRHBZ}: prints out the total vector density in the $xz$ plane limited to
$x>0$ because of axial symmetry. The first column denotes the coordinate $x$,
the second $z$, and the third column is the corresponding value of the total vector density.
\item {\tt DIRHBT}: prints out the total vector density in the
$xy$ ({\tt dirhb-xy.plo}), $xz$ ({\tt dirhb-xz.plo}) and $yz$ ({\tt dirhb-yz.plo}) planes;

\end{itemize}

\section{Acknowledgments}
This work was partly supported by the MZOS Project No. 1191005-1010, and the DFG cluster of excellence \textquotedblleft Origin and Structure of the Universe\textquotedblright\ (www.universe-cluster.de). T.N. acknowledges support from the Croatian Science Foundation.








\end{document}